\documentclass[onecollarge]{svjour2}

\usepackage{snapshot}

\usepackage{amsmath,amsfonts,amssymb}
\usepackage{verbatim}
\usepackage{graphicx}
\usepackage{pstricks}
\usepackage{appendix}
\usepackage{natbib}
\usepackage{feynmf}
\usepackage{float}

\usepackage{mathptmx}
\usepackage{latexsym}

\newcommand {\im}{\operatorname{Im}}

\newcommand{\D}[2]{\frac{\partial #1}{\partial #2}}

\newcommand{\Grr}{G_{\rho \rho}}
\newcommand{\GrrO}{G_{\rho \rho}^{(0)}}
\newcommand{\Grb}{G_{\rho B}}
\newcommand{\GrbO}{G_{\rho B}^{(0)}}
\newcommand{\Gbr}{G_{B \rho}}
\newcommand{\GbrO}{G_{B \rho}^{(0)}}

\newcommand{\rbs}{\rho \beta^2}

\numberwithin{equation}{section}

\journalname{Journal of Statistical Physics}

\begin{document}
\title{Power-law Decay and the Ergodic-Nonergodic Transition in Simple Fluids}
\author{Paul Spyridis \and Gene F. Mazenko}
\institute{ Paul Spyridis \and Gene F. Mazenko \at
           The James Franck Institute and the Department of Physics,
           The University of Chicago, 
           Chicago, Illinois. 60637, USA.
}

\date{}

\maketitle

\begin{abstract}
It is well known that mode coupling theory (MCT) leads to a two-step power-law time decay in
dense simple fluids. We show that much of the mathematical machinery used in the MCT analysis
can be taken over to the analysis of the systematic theory developed in the Fundamental Theory
of Statistical Particle Dynamics \cite{FTSPD}. We show how the power-law exponents can be 
computed in the second-order approximation where we treat hard-sphere fluids with statics
described by the Percus-Yevick solution.
\end{abstract}

\section{Introduction}

The long time dynamics of dense classical fluids is complicated. There is experimental
evidence of power-law decays as a function of time. One approach used to describe this
evolution is Mode-Coupling Theory (MCT). MCT has been an extremely controversial
theory. The problem is that there is no systematic derivation. MCT has been cobbled 
together piece by piece leading to its present form. There has been no systematic way 
of deriving, correcting, or extending the basic description.

Despite these criticisms MCT has led to a long time scenario which seems
to match a lot of the physical operations in a dense fluid. It has been argued by
\cite{AndreanovLandau} that MCT is the ``mean field'' theory governing glassy behavior 
and the basic results are more general than found in MCT.

The most contentious issue in judging MCT is the existence of an ergodic-nonergodic (ENE) transition. 
It is commonly believed that there are no ENE transitions in the classes of models studied.
However there is almost a transition. In the ideal MCT scenario there is an ENE 
transition at some critical packing fraction $\eta^*$. MCT implies a rather 
elaborate slow dynamics on either side of the critical density. There is, 
for $\eta < \eta^*$, a three step process \cite{KobReview}. The equilibrium 
density autocorrelation function first exhibits power-law decay
\[
\Grr(q, t) = \langle \rho_{-q}(0)\rho_q(t) \rangle_c \sim f_q + A_q t^{-a} \, ,
\]
followed by the von Schweidler decay
\[
\langle \rho_{-q}(0)\rho_q(t) \rangle_c \sim f_q - B_q t^{b} \, ,
\]
and finally exponential (possibly \emph{stretched}) decay
\[
\langle \rho_{-q}(0)\rho_q(t) \rangle_c \sim e^{-(t/t_0)^\beta} \, .
\]
A full analysis of the theory gives expressions for the exponents
\[
\frac{\Gamma(1 - a)^2}{\Gamma(1 - 2a)} = \lambda = \frac{\Gamma(1 + b)^2}{\Gamma(1 + 2b)} \, ,
\]
where $\lambda$ is a material dependent quantity. 

In FTSPD\cite{FTSPD} and SDENE\cite{SDENE} one of us introduced a new 
theory for the dynamics of collections of classical particles. It is shown
in \cite{SDENE} that at second order in perturbation theory one is led to
a description which has similarities with MCT.

In this paper we take the analysis of the kinetics beyond the analysis of 
the location of the transition. We explore the slow kinetics for the 
regimes bellow and above the transition. It was shown in \cite{SDENE, MMS} that the
density-density cumulant $G_{\rho \rho}(q, t)$ satisfies the 
kinetic equation
\begin{equation}\label{KineticEquation}
\frac{\partial G_{\rho \rho}}{\partial t}(q, t) 
  = -\bar{D}q^2 \bar{\rho} S(q)^{-1} G_{\rho \rho}(q,t)
      + \bar{D} q^2 \int_0^t ds \, \beta^2 \bar{\rho} \Sigma_{BB}(q, t-s) \dot{G}_{\rho \rho}(q, s)
\, ,
\end{equation}
where $\beta = 1/k_B T$ is the inverse temperature, $\bar{\rho}$ is the average density,
$\bar{D}$ is the diffusion coefficient, 
$S(q)$ is the static structure factor, and $\Sigma_{BB}$ is the 
B-B component of the self-energy matrix for the theory of the two
fields $\rho$ and $B$ where $B$ is a response field. The important
point is that $\Sigma_{BB}$ can be determined systematically in
perturbation theory and, at second order in the potential, is given by 
\begin{multline*}
\beta^2 \bar{\rho}^4 \Sigma_{BB}(q_1, t) 
  = -\frac{1}{2} \int \! \frac{d^3 k_2}{(2\pi)^3}\, d^3 k_3 \, dt' \, \delta(q_1 - k_2 - k_3)
    \, \delta(t - t')
    \bigg[ 1 + 2 \bar{K}_{13} \D{}{t} + 2\bar{K}_{12} \D{}{t'}
     + \bar{K}^2_{12}\D{^2}{{t'}^2}\\ + \bar{K}^2_{13} \D{^2}{t^2} 
     \left. + 2\bar{K}_{12}\bar{K}_{13} \D{}{t}\D{}{t'}\right] 
    \times \left[ G^{(0)}_{\rho \rho}(k_2, t) \tilde{G}_{\rho \rho}(k_3, t') + \bar{G}_{\rho \rho}(k_2, t)\bar{G}_{\rho \rho}(k_3, t') \right]
\end{multline*}
where $\bar{K}_{12} = (q_1 \cdot k_2) / q^2_1 k^2_2$, $\bar{K}_{13} = (q_1 \cdot k_3) / q^2_1 k^2_3$,
$G^{(0)}_{\rho \rho}$ is the non-interacting density autocorrelation function and 
$\tilde{G}_{\rho \rho}$ and $\bar{G}_{\rho \rho}$ are dressed propagators \cite{MMS}. Recently it 
has been suggested \cite{MCT-Test} that it is important to include information in 
addition to the static structure factor to make accurate quantitative predictions 
about glassy dynamics. In principle, a systematic extension of FTSPD to higher orders 
in the effective potential can take such data into account.

There are a number of differences between the MCT kernel and
$\Sigma_{BB}$ although the overall structure is similar. In
the FTSPD the self-energy has the basic structure represented
by the one-loop diagrams shown in Figure 1.
\begin{figure}
  \centering
  \begin{fmffile}{sigma1} 	
    \begin{fmfgraph*}(300, 50)
      \fmfpen{thick}
      \fmfleft{l} \fmfright{r}
      \fmfrpolyn{shaded}{A}{3}
      \fmfpolyn{shaded}{B}{3}
      \fmf{phantom}{l,A1}
      \fmf{phantom}{B1,r}
      \fmf{plain,left=.25,tension=.5}{A2,B2}
      \fmf{plain,right=.25,tension=.5}{A3,B3}
      \fmflabel{$1$}{A1}
      \fmflabel{$2$}{B1}
    \end{fmfgraph*}
  \end{fmffile}
  \caption{Generic one loop contribution to $\Sigma_{BB}(1,2)$.}
\end{figure}
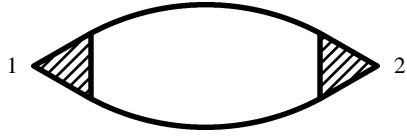
The triangles at the ends of the graph are the lowest order vertices.
We study here the cases where the vertex functions are given by the
non-interacting values. These vertices have a frequency dependence.
Between the vertices we have several sets of products of matrix 
propogators. An interesting feature is that all of these propagators
$G^{(0)}_{ij}$, $G_{ij}$, $\tilde{G}_{ij}$, $\bar{G}_{ij}$ satisfy similar 
fluctuation-dissipation theorems. For a full discussion see \cite{Newt1}.

It has been understood for some time that there are corrections 
to MCT that consist of higher numbers of loops. The two loop 
correction is of the general form
\begin{align*}
\begin{array}[t]{ccc}
\Sigma^{(3)}_{BB}(1,2) & \parbox{18pt}{\text{\huge$=$}} &
\parbox{100pt}{
  \begin{fmffile}{sigma2} 	
    \begin{fmfgraph*}(100, 47)
      \fmfpen{thick}
      \fmfleft{l} \fmfright{r}
      \fmfrpolyn{shaded}{A}{4}
      \fmfpolyn{shaded}{B}{4}
      \fmf{phantom,tension=10}{l,A1}
      \fmf{phantom,tension=10}{B1,r}
      \fmf{plain,left=.25,tension=.5}{A2,B2}
      \fmf{plain}{A3,B3}
      \fmf{plain,right=.25,tension=.5}{A4,B4}
      \fmflabel{$1$}{A1}
      \fmflabel{$2$}{B1}
    \end{fmfgraph*}
  \end{fmffile}}
\end{array}
\end{align*}
where now one has 4-point vertices and a product of these
dressed matrix propagators.

Our work here is to explore the long time solutions of the kinetic
equations. Near an ENE transition we can assume certain analytic forms
for the time dependence of $G_{\rho \rho}(q,t)$. For MCT a rather
elaborate mathematical machinery has been developed for performing
this asymptotic analysis \cite{G}. While none of this is new for MCT, we find it
useful to present these procedures in some detail. The results of an
analysis for MCT are well known; for the case of the FTSPD the results
are all new and comparable to the MCT. We begin with some background
information on FTSPD and MCT. We follow this by a review of the
asymptotic MCT analysis. Finally, we apply the analyis to FTSPD and
numerically determine the power-law and von Schweidler exponents
for hard spheres using the Percus-Yevick approximation for the
static structure factor. We compare these results to those of
the corresponding mode coupling theory.

\section{Reviews of FTSPD and MCT}

In this section we shall briefly review the origins of FTSPD and
MCT and how one can use them to construct kinetic equations like 
\eqref{KineticEquation} for correlation functions of interest.

\subsection*{FTSPD}

The Fundamental Theory of Statistical Particle Dynamics is a framework from which
mode-coupling-like kinetic equations can be derived, including systematic corrections
to them. In Smoluchowski dynamics, FTSPD is developed as a field theory whose fundamental
field is the position variable, $R_i(t)$. Starting with the equations of motion,
\[
\dot{R}_i (t) = D F_i(R) + \eta_i(t) \, 
\]
together with the variance of the Gaussian noise term
\[
\langle \eta_i(t) \eta_j(t') \rangle  = 2 k_B T D \delta_{ij} \delta(t - t') \, ,
\]
we obtain a generating functional for a path
\[
Z(H) = \langle e^{R_i \cdot H_i} \rangle = \int [d \eta_i] d^3 R_i^{(0)} P_0(R_i^{(0)})
       \exp \left\{- \frac{1}{4k_B T D} \int_{t_0}^\infty \eta_i^2 dt + R_i \cdot H_i \right\} \,
\]
where $R \cdot H$ is shorthand for
\[
\int_{t_0}^\infty R(t) \cdot H(t) \, dt \,
\]
and the integrated-over initial conditions are
\[
R_i(t_0) = R_i^{(0)} \, 
\]
with probability distribution given by $P_0(R_i^{(0)})$.
We can obtain the Martin-Siggia-Rose action for this system by exploiting the 
functional delta-function identity
\[
\int [d\eta_i] \, \delta(\dot{R}_i - DF_i - \eta_i) 
   = \int [dR_i] \, \det M_i \, \delta(\dot{R}_i - DF_i - \eta_i)
   = 1 \, ,
\]
where $\det M_i$ is the Jacobian
\[
\det M_i = \det \frac{\delta \eta_i}{\delta R_i} \, .
\]
Inserting this identity into the generating functional and using the
functional Fourier representation of the delta-function we obtain
\begin{align*}
Z(H) &= \int [d \eta_i] d^3 R_i^{(0)} P_0(R_i^{(0)})
         \exp \left\{- \frac{1}{4k_B T D} \int_{t_0}^\infty \eta_i^2 dt + R_i \cdot H_i \right\} \\
    &=  \int [d \eta_i] [dR_i] d^3 R_i^{(0)} P_0(R_i^{(0)}) \, 
        \det M_i \, \delta(\dot{R}_i - DF_i - \eta_i)
       \exp \left\{- \frac{1}{4k_B T D} \int_{t_0}^\infty \eta_i^2 dt + R_i \cdot H_i \right\} \\
    &= \int [d \eta_i] [dR_i] [d\hat{R}_i] d^3 R_i^{(0)} P_0(R_i^{(0)}) \, \det M_i \, 
       \exp \left\{ i \int \hat{R_i}(\dot{R}_i - DF_i - \eta_i) dt \right\}
       \exp \left\{- \frac{1}{4k_B T D} \int_{t_0}^\infty \eta_i^2 dt + R_i \cdot H_i \right\} \, .
\end{align*}
We now combine the arguments of the two exponentials into
a single integral:
\begin{align*}
Z(H) &= \int [d \eta_i][dR_i] [d\hat{R}_i] d^3 R_i^{(0)} P_0(R_i^{(0)}) \, \det M_i \, \exp
\left\{\int_{t_0}^\infty \left(i\hat{R_i}(\dot{R}_i-DF_i-\eta_i)-\frac{\eta_i^2}{4k_B T D} \right)dt + R_i \cdot H_i \right\}\\
&= \int [d \eta_i] [dR_i] [d\hat{R}_i] d^3 R_i^{(0)} P_0(R_i^{(0)}) \, \det M_i \, \exp
\left\{\int_{t_0}^\infty \left( i\hat{R_i}(\dot{R}_i-DF_i)-\frac{\eta_i^2 + 4ik_BTD\hat{R}_i \eta_i}{4k_B T D} \right) dt + R_i \cdot H_i \right\} \\
&= \int [d \eta_i] [dR_i] [d\hat{R}_i] d^3 R_i^{(0)} P_0(R_i^{(0)}) \, \det M_i \, \\
&\hspace{5em} \times \exp
\left\{\int_{t_0}^\infty \left( i\hat{R_i}(\dot{R}_i-DF_i)-\frac{(\eta_i + 2ik_BTD\hat{R}_i)^2}{4k_B T D} - \hat{R}_ik_BTD\hat{R}_i \right)\, dt + R_i \cdot H_i \right\} \, .
\end{align*}
Now all other variables are independent of $\eta_i$, so we can perform the Gaussian integral
and absorb the constant into the measure. We obtain
\[
Z(H) = \int [dR_i] [d\hat{R}_i] d^3 R_i^{(0)} P_0(R_i^{(0)}) \, \det M_i \, \exp
\left\{ -\int \hat{R}_ik_BTD\hat{R}_i - i\hat{R_i}(\dot{R}_i-DF_i) \, dt + R_i \cdot H_i \right\}
\, .
\]
We use this to build the generating functional for all the paths:
\[
Z_N(H) = \int \prod_{i = 1}^N [dR_i] [d\hat{R}_i] d^3 R_i^{(0)} P_0(R_i^{(0)}) \, \exp
\left\{ -\int dt \sum_{i = 1}^N \, [\hat{R}_ik_BTD\hat{R}_i - i\hat{R_i}(\dot{R}_i-DF_i) - A^i_J] + \phi \cdot H \right\}
\, ,
\]
where we have rewritten the Jacobians as $A^i_J = \ln \det{M_i}$ and introduced the symbol $\phi$
to represent the fields (constructed from $R$, $\hat{R}$) that we wish to couple to $H$. 
A more complete derivation with a treatment of the Jacobian is  
available in \cite{FTSPD}. From the generating functional of the paths we obtain the
grand canonical partition function
\[
Z_T[H] = \sum_{N = 0}^\infty \frac{z^N}{N!} Z_N(H)
\]
from which we construct the generator of cumulants
\[
W[H] = \log Z_T[H] \, .
\]
The source fields $H_\rho$ and $H_B$ are coupled to the density
\[
\rho(1) = \sum_i \delta(x_1 - R_i(t_1)) \, ,
\]
and response field
\[
B(1) = D \sum_i \left\{ \hat{R}_i \nabla_1 + \theta(0)\nabla_1^2 \right\} \delta(x_1 - R_i(t_1)) \, ,
\]
respectively. The value of $\theta(0)$ is chosen based upon the discretization scheme we
adopt for the path integral. The standard Stratonovich discretization\footnote{Using the Ito 
prescription is equivalent to making $\theta(0) = 0$, resulting in a trivial Jacobian. The 
Stratonovich prescription is usually preferable since the ordinary chain rule of calculus 
remains unmodified when dealing with multiplicative noise. Either way, in FTSPD the Fokker-Planck
equation and all correlation functions resulting from the path integral are independent of 
the choice of discretization since the noise is not multiplicative.} 
results in $\theta(0) = 1/2$. 
We generate cumulants by functional differentiation:
\[
G_{\rho...\rho B...B}(1,...,\ell, \ell+1, ..., n) 
    = \frac{\delta}{\delta H_\rho} \cdots \frac{\delta}{\delta H_\rho} 
        \frac{\delta}{\delta H_B} \cdots  \frac{\delta}{\delta H_B} W[H] \bigg|_{H_\rho = H_B = 0}
\, \, .
\]
From the Legendre transform of $W[H]$ one generates the irreducible vertices
\[
\Gamma_{ij...k} = \frac{\delta}{\delta G_i} \frac{\delta}{\delta G_j} \cdots
                 \frac{\delta}{\delta G_k} W[G] \, .
\]
The two-point vertices satisfy the Dyson equation
\[
\sum_k \Gamma_{ik} G_{kj} = \delta_{ij} \, .
\]
Expanding the vertex to second order in the potential we have
$\Gamma_{ik} = \gamma_{ik}^{(0)} + \gamma_{ik}^{(1)} - \Sigma_{ik}$.
Plugging into the $B$-$\rho$ component of Dyson's equation 
we construct the kinetic equation
\[
\frac{\partial G_{\rho \rho}}{\partial t}(q, t) 
  = -\bar{D}q^2 \bar{\rho} S(q)^{-1} G_{\rho \rho}(q,t)
      + \bar{D} q^2 \int_0^t ds \, \beta^2 \bar{\rho} \Sigma_{BB}(q, t-s) \dot{G}_{\rho \rho}(q, s)
\, ,
\]
where $\bar{D} = k_B T D$ and $G_{\rho \rho}(q, 0) = S(q)$. It is shown in \cite{SDENE} that 
approximations for $\Sigma_{BB}$ can be generated as a power series in an effective 
interaction potential.

\subsection*{MCT}

The most common path to Mode Coupling Theory is through the projection operator formalism
of Zwanzig and Mori. We briefly summarize the development given in \cite{ReichmanReview}.
We start from the fundamental relationship between a function of phase-space variables $A(r,p)$
and the Hamiltonian $\mathcal{H} = \sum_i p_i^2/2m + \frac{1}{2}\sum_{i \neq j}V(r_{ij})$:
\begin{equation} \label{A-Evolution}
\frac{dA}{dt} = \{A, \mathcal{H} \} \equiv i \mathcal{L} A(t) \, ,
\end{equation}
where we have used the Poisson bracket
\[
\{A, B\} = \sum_i \left( \frac{\partial A}{\partial r_i} \frac{\partial B}{\partial p_i}
                        -\frac{\partial A}{\partial p_i} \frac{\partial B}{\partial r_i} 
                  \right) \, .
\]
This equation defines the Liouville operator $\mathcal{L}$:
\[
i \mathcal{L} = \sum_i \frac{p_i}{m} \frac{\partial}{\partial r_i} 
                - \sum_{i \neq j} \frac{\partial V}{\partial r_i} \frac{\partial}{\partial p_i}
\, .
\]
It follows that we may formally express the time evolution of $A$ as 
\[
A(t) = e^{i\mathcal{L}t} A(0) \, .
\]
The next step in the analysis is to identify $A$ with a set of functions
that we are interested in tracking. These are referred to as ``slow variables.''
For the derivation of the basic MCT equation these are the density fluctuations 
and longitudinal current: $A(t) = (\delta \rho_k(t), j_k^{\text{L}}(t))$. We also 
define an inner product
\[
(A, B) = \langle A^\dagger B \rangle \, ,
\]
which allows us to define the projection operator
\footnote{The projection operator retains the same form when $A$ is generalized 
to a set of functions. In this case, $(A, A)$ refers to the matrix $A_{ij} = (A_i, A_j)$
and $\mathcal{P} B = \sum_{i, j} {(A, A)^{-1}}_{ij} (A_j, B) \, A_i$.}
\[
\mathcal{P} \equiv A(A, ...)(A, A)^{-1} \, .
\]
Note that $A = A(0)$ above. The projections perpendicular to the slow variables 
are referred to as ``fast variables.'' Separating the left-hand side of (\ref{A-Evolution}) 
into components parallel and perpendicular to $A(0)$ allows us to express the 
time evolution as
\[
\frac{dA}{dt} = i \Omega A(t) + \int_0^t e^{i \mathcal{L}(t - \tau)} i \mathcal{PL}f(\tau)\, d\tau
                + f(t) \, ,
\]
where
\[
i \Omega = (A, i \mathcal{L} A)(A, A)^{-1}
\]
and 
\[
f(t) = e^{i(1 - \mathcal{P})\mathcal{L}t} i (1 - \mathcal{P}) \mathcal{L} A \, .
\]
Exploiting the orthogonality between $f(t)$ and $A(0)$ allows us to write
\[
\frac{dA}{dt} = i \Omega A - \int_0^t K(t - \tau) A(\tau) \, d\tau + f(t) \, ,
\]
where the kernel is given by
\[
K(t) = (f(0), f(t))(A(0), A(0))^{-1} \, .
\]
Introducing the correlation matrix $C(t) = \langle A^\dagger(0) A(t) \rangle = (A(0), A(t))$
we have
\[
\frac{dC}{dt} = i \Omega C(t) - \int_0^t d\tau \, K(t - \tau) C(\tau) \, .
\]
Specializing to our choice of slow variables, $A = (\delta \rho_k(t), j_k^{\text{L}}(t))$, we obtain
\begin{equation}\label{MCT-Exact}
\frac{\partial^2 \Grr(q,t)}{\partial t^2} + \frac{q^2 k_B T}{m S(q)} \Grr(q, t) 
   + \frac{m}{Nk_BT} \int_0^t d\tau \, \langle R_{-q}(0) R_q(t - \tau)\rangle \dot{G}_{\rho \rho}(q, \tau)
= 0 \, 
\end{equation}
from the $\rho$-$j$ component of our correlation matrix, where
\[
\Grr(q, t) = \langle \delta \rho_{-q}(0) \delta \rho_q(t) \rangle
\]
is the density-density cumulant \footnote{It should be noted that the convention 
used in this section (and most of the MCT literature) for the cumulants is different from 
the convention we have employed in the discussion of FTSPD: 
$\Grr^{FT}(q, t) = \lim \, V^{-1} \langle \delta \rho_{-q}(0) \delta \rho_q(t) \rangle$ 
while $\Grr^{MCT}(q, t) = \lim \, N^{-1} \langle \delta \rho_{-q}(0) \delta \rho_q(t) \rangle$,
where $V$ and $N$ are the volume and number of particles, respectively.
Accordingly the static structure factors are related in the same way: $S^{FT}(q) = \rho S^{MCT}(q)$. 
We continue to use the MCT convention when we discuss the MCT kinetic equation below, 
and switch back to the original convention when we discuss the FTSPD kinetic equation.}
and
\[
R_q = \frac{dj_q^{\text{L}}}{dt} - i \frac{q k_B T}{m S(q)} \delta \rho_q \, .
\]
So far all of these manipulations have been exact. However, computing with (\ref{MCT-Exact}) 
is not practical without simplifying the kernel $K(t) = \langle R_{-q}(0) R_q(t)\rangle$. 
This is the point where the uncontrolled approximations of MCT are introduced. They 
consist of:
\begin{itemize}
\item Replacing the time evolution operator 
  $e^{i(1 - \mathcal{P})\mathcal{L}t} \to \mathcal{P}_2 e^{i \mathcal{L}t} \mathcal{P}_2$, where
  $\mathcal{P}_2$ projects onto the dominant ``slow'' mode of $R_q$.
\item Gaussian and convolution approximations for the correlation functions which allow them 
  to be factorized into simpler correlation functions proportional to the static structure
  factor.
\end{itemize}
In contrast to the FTSPD, it is not clear how one could go about systematically improving the MCT 
approximations for the memory function kernel. 

\section{Review of MCT Asymptotic Analysis}

These approximations define mode coupling theory and lead to the following 
equation\footnote{We are presenting the Newtonian MCT equation. For Smoluchowski 
dynamics only one time derivative appears.} for the density autocorrelation 
function, $\Grr(q, t)$:
\[
\frac{\partial^2}{\partial t^2} \Grr(q,t) + \Omega^2(q) \Grr(q,t) 
+ \Omega^2(q) S(q) \int_0^t \; dt' M(q, t - t') \frac{\partial}{\partial t'} 
\Grr(q, t')  = 0\;, 
\]
where 
\begin{align*}
\Omega^2(q) &= \frac{q^2 k_B T}{m S(q)} \\
M(q, t) &= \frac{1}{2} \int \; \frac{d^3 k}{(2 \pi)^3} V^2(\vec{q}, \vec{k}) 
                                 \Grr(k, t) \Grr(|\vec{q} - \vec{k}|, t) \\
V^2(\vec{q}, \vec{k}) &= \frac{\rho}{q^2} \left\{ \hat{q}\cdot \vec{k} c(k) + 
         \hat{q} \cdot (\vec{q} - \vec{k}) c(|\vec{q} - \vec{k}|)  \right\}^2 \\
\rho c(k) &=  1 - \frac{1}{S(k)}
\end{align*}
The important point is that $M$ is
a non-linear function of the density autocorrelation function. $V(q, k)$ is the 
vertex in the theory. We may rewrite the equation in terms of the normalized density 
autocorrelation function $\phi(k, t)$ which is related to the 
autocorrelation function by $\Grr(q,t) = S(q) \phi(q, t)$. The kinetic equation can
then be written as
\begin{equation}\label{MCT-Kinetic}
\frac{\partial^2}{\partial t^2} \phi(q,t) + \Omega^2(q) \phi(q,t) + 
\Omega^2(q) \int_0^t \; dt' M'(q, t - t') \frac{\partial}{\partial t'} 
\phi(q, t') = 0 \;.
\end{equation}
The corresponding kernel $M'(q,t)$ is given by
\begin{align*}
M'(q,t) &= \frac{1}{2} \int \; \frac{d^3 k}{(2 \pi)^3} 
                   V^2(\vec{q}, \vec{k}) \phi(k, t) \phi(|\vec{q} - \vec{k}|, t) \\
V^2(\vec{q}, \vec{k}) &= \frac{\rho}{q^2} S(q) S(k) S(|\vec{q} - \vec{k}|)
\left\{ \hat{q}\cdot \vec{k} c(k) + \hat{q} \cdot (\vec{q} - \vec{k}) 
c(|\vec{q} - \vec{k}|)  \right\}^2 \; .
\end{align*}
The only relevant parameter for hard spheres is the number density, $\rho$. 
We are interested in extracting the long-time slow dynamics for this model. 
In the following sections all numerical calculations and plots are found
using the Percus-Yevick approximation for the hard sphere sturcture factor.
It is useful to work using the Laplace 
transform\footnote{Our conventions are defined in Appendix B.} 
which diagonalizes \eqref{MCT-Kinetic} \cite{BGS, KM}. 
This leads to
\[
z^2 \phi(q, z) - z + \Omega^2 \phi(q, z) + 
\Omega^2 \mathcal{L}[ M(q, t)](z \phi(q, z) - 1) = 0 \; ,
\]
where we have applied the initial conditions $\phi(q, 0) = 1$ and 
$\dot{\phi}(q, 0) = 0$. Rearranging terms we have
\begin{equation}\label{MCT}
\frac{\Omega^2 \phi(q, z)}{1 - z \phi(q, z)} = 
\Omega^2 \mathcal{L}[ M(q, t)] + z \; .
\end{equation}

\subsection{Heuristic Analysis of MCT: Time Dependence at Transition}\label{Heuristic}

\begin{figure}[h]
  \centering
  \input{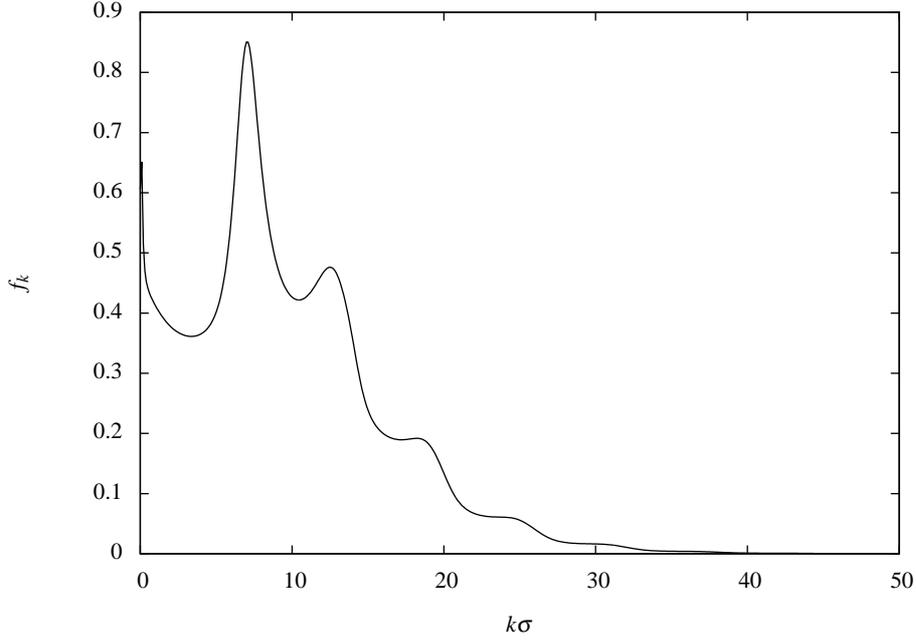}
  \caption{This plot shows $f_k^{\text{MCT}}$, the non-ergodicity parameter, at the
    critical density $\eta_{\text{MCT}} = 0.515$.}
  \label{f_k-plot}
\end{figure}

\begin{figure}[h]
  \centering
  \input{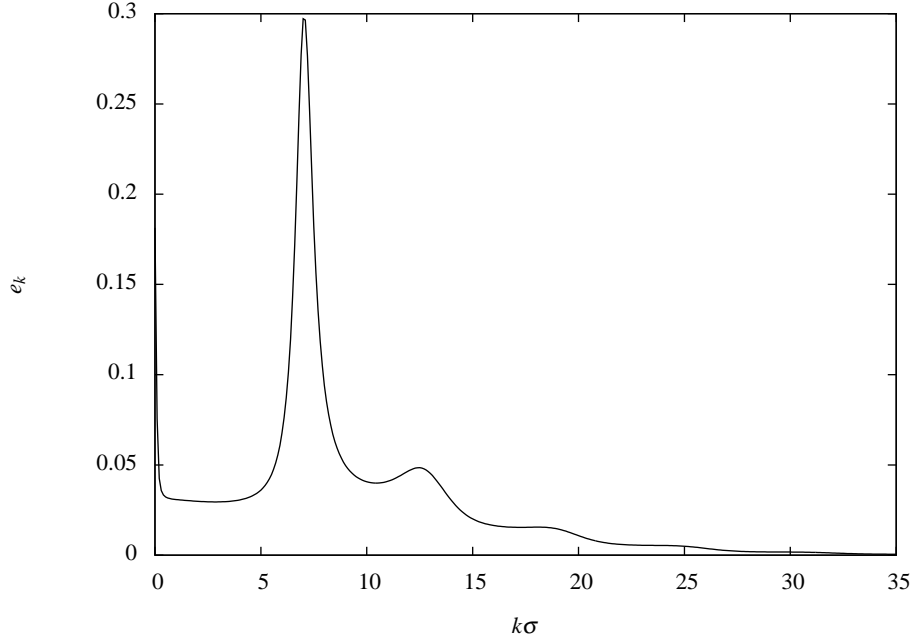}
  \caption{This plot shows $e_k^{\text{MCT}}$, the right eigenvector of $C_{qk}^{\text{MCT}}$, at 
    the critical density $\eta_{\text{MCT}} = 0.515$. The eigenvector is normalized
    according to the convention $\sum_k e_k e_k = 1$. }
  \label{fig:e_k}
\end{figure}

\begin{figure}[h]
  \centering
  \input{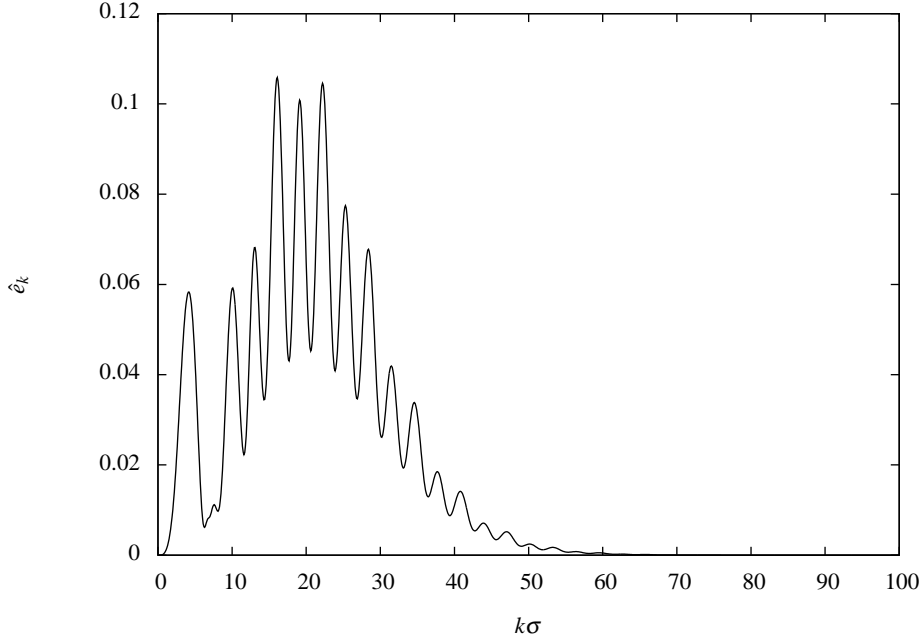}
  \caption{This plot shows $\hat{e}_k^{\text{MCT}}$, the left eigenvector of $C_{qk}^{\text{MCT}}$, at 
    the critical density $\eta_{\text{MCT}} = 0.515$. The eigenvector is normalized
    according to the convention $\sum_k \hat{e}_k \hat{e}_k = 1$. }
  \label{fig:he_k}
\end{figure}

In the long-time regime we assume $\phi(q, z) \to \infty$ as $|z| \to 0$. 
In this regime ($|z| \ll 1$) we may ignore the $z$ on the right 
hand side of (\ref{MCT}) and thereby cancel the factors of $\Omega$. This gives 
us\footnote{Note that in MCT both Newtonian dynamics and Smoluchowski
dynamics satisfy the same asymptotic equation.}
\begin{equation}\label{MCT2}
\frac{ \phi(q, z)}{1 - z \phi(q, z)} =  \mathcal{L}[ M(q, t)] \; .
\end{equation}
Next we apply the ansatz $\phi(q, t) = f(q) + (1 - f(q))^2 \psi_q(t)$ where
$\psi_q(t)$ is assumed small and $f(q)$ is a non-ergodicity parameter. The Laplace 
transform of the ansatz is given by
\[
\phi(q, z) = \frac{f(q)}{z} + (1 - f(q))^2\psi_q(z) \; .
\]
We now plug this into (\ref{MCT2}) and expand the left hand side to second order in $\psi_q(z)$:
\[
\frac{1}{1 - f_q} \left( \frac{f_q}{z} + (1 - f_q)\psi_q(z) + z 
(1 - f_q)^2 \psi_q^2(z) \right) = \mathcal{L}[ M(q, t)] \; .
\]
To apply the ansatz to the right hand side note that
\begin{align*}
\phi(k,t) \phi(|\vec{q} - \vec{k}|, t) & = \{f_k + (1 - f_k)^2\psi_k (t)\}\{f_{q - k} + 
(1 - f_{q-k})^2\psi_{q-k} (t) \} \\
			       &= f_k f_{q - k} + 
f_k ( 1 - f_{q - k})^2 \psi_q(t) + f_{q - k} (1 - f_k)^2 \psi_k (t) + 
(1 - f_k)^2(1 - f_{q - k})^2  \psi_k(t) \psi_{q - k}(t) \; .
\end{align*}
The two middle terms above may be combined into a single term in the integral. 
We now employ the notation of \cite{G} to make the following more compact. Let
\begin{align}
F_q[f_k] &=  \frac{\rho S(q)}{2q^2}  \int \frac{d^3 k}{(2 \pi)^3} 
V^2(\vec{q}, \vec{k}) f_k f_{q - k} \notag \\
C_{q} &= \rho \frac{\delta F_q[f_k]}{\delta \rho} \notag \\
C_{qk} &= \frac{\delta F_q[f_k]}{\delta f_k} (1 - f_k)^2 \label{DefCqk}\\
C_{qkp} &= \frac{1}{2}\frac{\delta^2 F_q}{\delta f_k \delta f_p} 
(1 - f_k)^2 (1 - f_p)^2 \, . \notag
\end{align}
We obtain
\begin{equation}\label{MCT3}
\frac{1}{z}\left( \frac{f_q}{1 - f_q} -  F_q[f_k] \right) + \left( \psi_q(z) -  C_{qk} \psi_k(z) \right) + z(1 - f_q)\psi_q^2(z) - C_{qkp} \mathcal{L}[\psi_k(t) \psi_p(t)] = 0 \;
\end{equation}
where it is understood that we are integrating over repeated indices:
$C_{qk} f_k \equiv \int d^3 \vec{k} \, C_{qk}f_k / (2 \pi)^3$, etc.
This expansion yields three conditions for the ideal metastable state. 
The first one, given by
\[
\frac{f_q}{1 - f_q} =  F_q[f_k] \; 
\]
determines $f_q$. For hard spheres, this equation can be solved by iteration. 
A plot of $f_q$ using the Percus-Yevick approximation for the hard-sphere structure 
factor is available in Figure \ref{f_k-plot}.
The second and third conditions can be found by considering a further expansion of 
$\psi_q(z) = \psi_q^{(1)} + \psi_q^{(2)}$, with $\psi_q^{(2)}/\psi_q^{(1)} \to 0$ 
as $z \to 0$. We plug the expansion into our equation and match orders. 
To the order of $\psi_q^{(1)}$, we obtain
\begin{equation}\label{eigen}
\psi_q^{(1)}(z) = C_{qk} \psi_k^{(1)}(z) \; ,
\end{equation}
where $C_{qk}$ is defined by \eqref{DefCqk}. This equation determines the spatial 
dependence of $\psi_q^{(1)}(z)$. We can solve this equation for $\psi_q^{(1)}(z)$ 
by finding an eigenvector of the operator 
\[
C_{qk} = \frac{ \rho S(q)}{q^2}  
V^2(\vec{q}, \vec{k}) f_{q - k} (1 - f_k)^2
\]
with unit eigenvalue. This eigenvector is found at the critical density \cite{EigenProof}. 
Below the critical density (\ref{eigen}) cannot support a nontrivial solution. Above 
the critical density the eigenvalue moduli are all less than one. Denote the right 
eigenvector associated with the unit eigenvalue by $e_k$, and the corresponding
left eigenvector by $\hat{e}_k$. Furthermore, impose the conditions 
$\sum_k \hat{e}_k e_k = 1$ and $\sum_k \hat{e}_k e_k e_k (1 - f_k)  = 1$. 
With this notation we have
\[
\psi_q^{(1)} = A e_q \phi_\nu (z) \; ,
\]
with $A$ a thus far undetermined constant and $\phi_\nu(z)$ a function of $z$ 
only. Plots of $e_k$ and $\hat{e}_k$ for the hard sphere system are available in 
Figures \ref{fig:e_k} and \ref{fig:he_k}, respectively.
Next, we match the second order terms in (\ref{MCT3}). We obtain
\[
(\delta_{qk} - C_{qk})\phi_k^{(2)} = -(1 - f_q)z [\psi_q^{(1)}]^2 + 
C_{qkp} \mathcal{L}[ \psi_k^{(1)} \psi_p^{(1)}] \; .
\]
Now multiply both sides by the left eigenvector, $\hat{e}_q$. We obtain
\begin{align*}
0 &= -\hat{e}_q (1 - f_q)z [\psi_q^{(1)}]^2 + \hat{e}_q C_{qkp} \mathcal{L} [\psi_k^{(1)} \psi_p^{(1)}] \\
   &= -\hat{e}_q (1 - f_q)z A^2 e_q e_q \phi_\nu^2(z) + A^2 \hat{e}_q C_{qkp} e_k e_p \mathcal{L} [ \phi_\nu^2(t)]\\
   &= -A^2 z \phi_\nu^2(z) + A^2 \lambda \mathcal{L} [ \phi_\nu^2(t)] \; ,
\end{align*}
where we have used the condition  $\sum_k \hat{e}_k e_k e_k (1 - f_k)  = 1$ and
$\lambda = \hat{e}_q C_{qkp} e_k e_p$ in the third line. This gives us an 
equation for the $z$-dependent part of $\psi^{(1)}$:
\begin{equation}\label{MCT-Lambda}
z \phi_\nu^2(z) = \lambda \mathcal{L} [ \phi_\nu^2(t)] \; .
\end{equation}
Using\footnote{The amplitude of the power law is absorbed in the parameter $A$.} 
$\phi_\nu(t) = t^{-a}$, which implies $\phi_\nu(z) = \frac{\Gamma(1 - a)}{z^{1 - a}}$, 
we obtain a quantitative expression for the power law index
\[
z \frac{\Gamma^2(1 - a)}{z^{2 - 2a}} = \lambda \frac{\Gamma(1 - 2a)}{z^{1 - 2a}} \;.
\]
This implies
\[
\frac{\Gamma^2(1 - a)}{\Gamma(1 - 2a)} =  \lambda =  \hat{e}_q C_{qkp} e_k e_p \; ,
\]
where $\lambda$ has to be determined numerically. This analysis only applies near the 
transition density. The assumption of $\psi = \psi^{(1)} + \psi^{(2)}$ and matching the 
orders might fail in other circumstances. A question remaining in this analysis is the
nature of the separation of $\psi_q(z)$ into first and second order terms.

\subsection{Ergodic-Nonergodic Transition}\label{ENE-Transition}

Having demonstrated the approach corresponding to slow power-law decay, we
can push harder for a fuller description at the transition as well as above
and below. We take first the case of working at the transition.
To see if the mode-coupling approach supports an ergodic-nonergodic transition,
we write $\phi(q, t) = f_q + g_q(t)$, with $g_q(t) \to 0$ as $t \to \infty$.
This implies $\phi(q, z) = f_q/z + g_q(z)$, with $g_q(z)$ less singular
than $1/z$. Let us make this substitution in \eqref{MCT}:
\[
\frac{\Omega^2 (f_q/z + g_q(z))}{1 - z(f_q/z + g_q(z))} = 
\Omega^2 \mathcal{L}[ M(q, t)] + z \; .
\]
Ignoring terms less singular than $1/z$ in the long time limit 
($z \to 0$) we obtain
\begin{equation}\label{iter}
\frac{f_q}{1 - f_q} = F_q(f_k) \; .
\end{equation}
Therefore, we find that there is a non-ergodic state when the above equation
supports a nontrivial solution for $f_q$. Solving this equation by iteration
reveals that there is indeed an ergodic/non-ergodic transition for
hard spheres at $\eta \approx 0.52$, using the Percus-Yevick approximation
for the static structure factor.

\begin{figure}[h]
  \centering
  \input{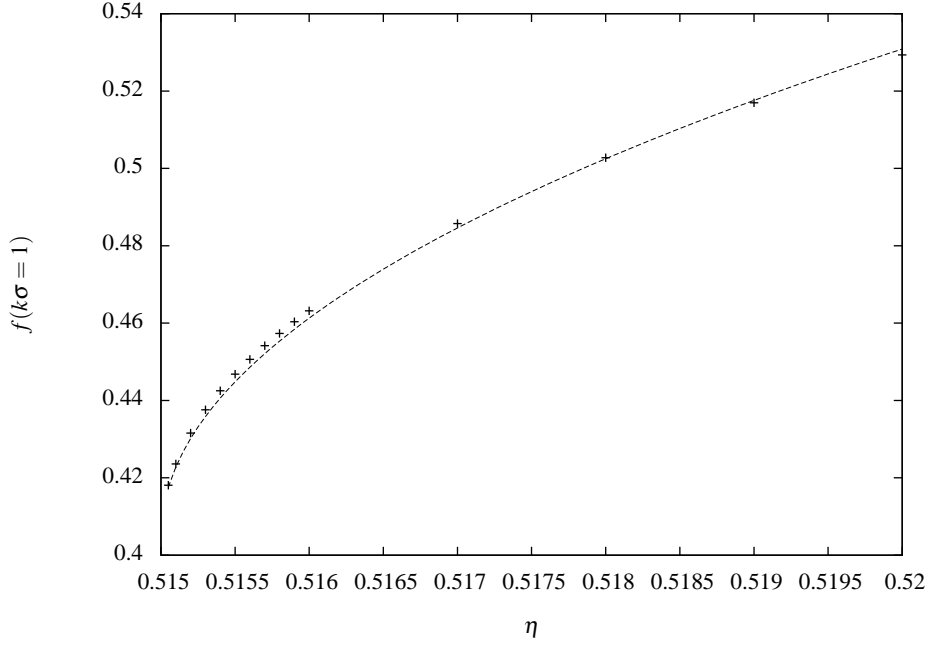}
  \caption{This plot shows $f^{\text{MCT}}(k \sigma = 1.0)$, the non-ergodicity parameter at
    wavenumber $k \sigma = 1.0$, as a function of the packing fraction. The 
    dashed line shows a visual fit of $A\sqrt{\epsilon} +  B$ to the data. This
    confirms our calculation of $g$ in Section \ref{ENE-Transition}.}
  \label{fig:f_k-scale}
\end{figure}

Next let us determine how the non-ergodicity parameter changes when we perturb
the density: $\rho \to \rho + \delta \rho$. Write $f_q = f_q^{\rho} + (1 - f_q^{\rho})^2 g_q$, where
$g_q$ is ``small'', and substitute into the left hand side of \eqref{iter}:
\[
\frac{f_q}{1 - f_q} = \frac{f_q^{\rho} + (1 - f_q^{\rho})^2 g_q}{(1 - f_q^{\rho})(1 + 
  (1 - f_q^{\rho})g_q)} = \frac{f_q^{\rho}}{1 - f_q^{\rho}} + g_q + (1 - f_q^{\rho})g_q^2 + O(g^3)
\]
The right hand side of \eqref{iter} expands to
\[
F_q^{\rho + \delta \rho} (f_k^{\rho + \delta \rho}) = F_q^{\rho} 
    + \frac{\delta F_q^{\rho}}{\delta {\rho}} \delta {\rho}
    + \frac{\delta F_q^{\rho}}{\delta f_k} \delta f_k 
    + \frac{1}{2}\frac{\delta^2 F_q^{\rho}}{\delta f_k \delta f_p} \delta f_k \delta f_p 
    + O(g^3, g\rho, \rho^2)\, ,
\]
where $\delta f_k = (1 - f_k)^2 g_k$ and $F_q^{\rho}$ is evaluated at $f_k^{\rho}$. Using
notation introduced in Section \ref{Heuristic} we obtain
\[
F_q^{\rho + \delta \rho} (f_k^{\rho + \delta \rho}) = 
          F_q^{\rho} + C_q \epsilon + C_{qk} g_k + C_{qkp} g_k g_p + O(g^3, g\rho, \rho^2) \, .
\]
Combining both sides we obtain an equation for $g_q$:
\[
g_q + (1 - f_q^{\rho})g_q^2 = 
     C_q \epsilon + C_{qk} g_k + C_{qkp} g_k g_p + O(g^3, g\rho, \rho^2) \, .
\]
Rewriting in a more suggestive form,
\begin{equation}\label{eigen-extra}
(\delta_{qk} - C_{qk})g_k = 
          -(1 - f_q^{\rho})g_q^2 + C_q \epsilon + C_{qkp} g_k g_p + O(g^3, g\rho, \rho^2)  
\end{equation}
At the transition density, $\rho = \rho_c$, $C_{qk}$ has unit eigenvalue. Let us
expand $g_k = g_k^{(1)} + g_k^{(2)}$ and assume that $g_k^{(1)}$ is the unit
eigenvalue of $C_{qk}$. In that case, we can write $g_k^{(1)} = g e_k$, with
$g$ a thus far undetermined small parameter. Plugging this form into 
\eqref{eigen-extra} allows us to relate $g$ and the separation parameter
$\epsilon = (\rho - \rho_c)/\rho_c$. Expanding to second order in $g$, we obtain
\[
(\delta_{qk} - C_{qk})g_k^{(2)} = 
          -(1 - f_q^{\rho})(g_q^{(1)})^2 + C_q \epsilon + C_{qkp} g_k^{(1)} g_p^{(1)} 
          + O(g^3, g\rho, \rho^2) \, .
\]
Multiplying both sides by the left eigenvector of $C_{qk}$ to eliminate
$g_k^{(2)}$ yields
\[
0 = -g^2 \hat{e}_q (1 - f_q^{\rho})e_q^2 + \hat{e}_q C_q \epsilon 
     + g^2 \hat{e}_q C_{qkp} e_k e_p + O(g^3, g\rho, \rho^2) \, .
\]
Recalling the convention for the eigenvectors, $\hat{e}_q (1 - f_q^{\rho})e_q^2 = 1$,
and the definition of $\lambda = \hat{e}_q C_{qkp} e_k e_p$, we find
\[
g^2 = \frac{\hat{e}_q C_q \epsilon}{1 - \lambda} \, .
\]
We also know that $C_q$ is positive, since to leading order in the separation
parameter $F_q^{\rho + \delta \rho} = (1 + \epsilon)F_q^{\rho}$. Therefore our assumptions
are consistent near the transition density as long as $\lambda < 1$ and the 
small parameter in $g_k^{(1)}$ scales as  $\sqrt{\epsilon}$. Putting all of this
together, the non-ergodicity parameter near the transition density can be
written as
\[
f_q = f_q^c + g (1 - f_q)^2 e_q \, .
\]
A plot demonstrating the proper scaling behavior is shown in Figure \ref{fig:f_k-scale}. 
A similar analysis implies that for the largest eigenvalue $E_0$ of $C_{qk}$,
assuming the eigenvalue is one at the critical density, 
$E_0 = 1 - 2g(1 - \lambda)$, agreeing with the statements in \cite{BGL}. A
plot of the maximum eigenvalue of $C_{qk}$ as a function of the packing fraction
is shown in Figure \ref{fig:C_{qk}-max-eigen}.
\begin{figure}[h]
  \centering
  \input{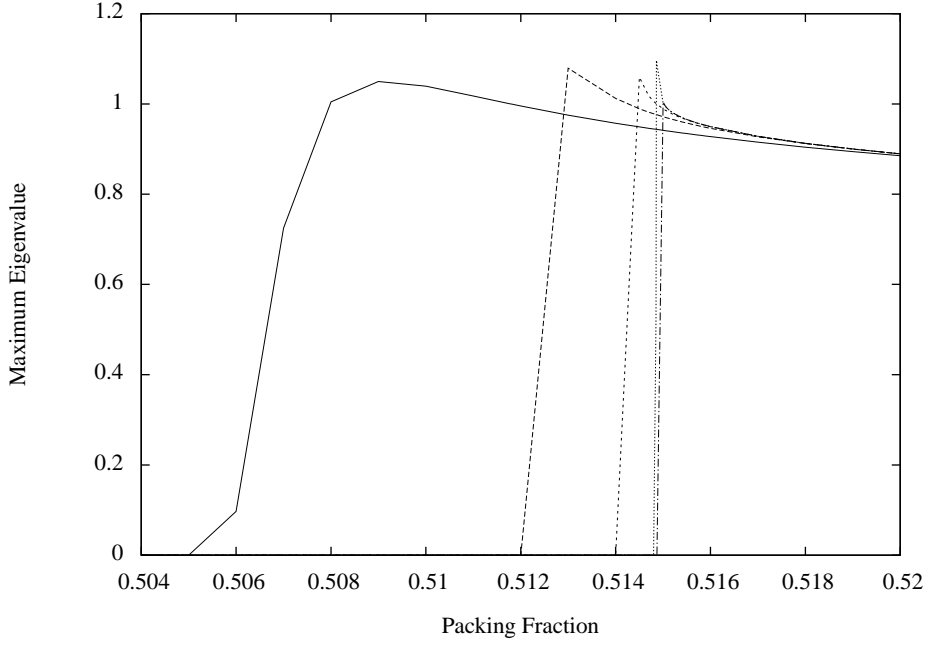}
  \caption{This plot shows the maximum eigenvalue of $C_{qk}^{\text{MCT}}$ as a function
    of density. The lines are drawn at 30, 60, 120, 240, and 270 iterations. 
    We see the lines converging as the number of iterations increases.}
  \label{fig:C_{qk}-max-eigen}
\end{figure}

\subsection{Time regimes}

In this section we will analyze the asymptotic long-time behavior of
$\phi(q, t)$. First we need to find out what happens at the critical
density so we can match the other solutions to it in the 
limit $\epsilon \to 0$. The analysis of $\phi(q, t)$ at the critical
density is essentially the same as in Section \ref{Heuristic}, but we 
include it here for completeness. In the non-ergodic (glass) case
we have only one long time to consider. In the ergodic (liquid) case
there are two time regimes: the power law decay as in the solid case
followed by the von Schweidler decay. To analyze these regimes we rescale
the time $t$ by introducing a large constant $\omega_c$ such that $\tau = \omega_c \, t$.
In this rescaling the power law decay corresponds to $\tau \ll 1$ and
the von Schweidler decay to $\tau \gg 1$.

\subsubsection{Critical Density}

Here we are interested in the asymptotic behavior of $\phi(q, t)$ at large
times. First we consider what happens at the critical density.  
As in Section \ref{Heuristic} we apply the ansatz 
$\phi(q, t) = f(q) + (1 - f(q))^2 \psi_q(t)$.
We obtain
\begin{equation}\label{MCT4}
\frac{1}{z}\left( \frac{f_q}{1 - f_q} -  F_q[f_k] \right) + \left( \psi_q(z) -  C_{qk} \psi_k(z) \right) + z(1 - f_q)\psi_q^2(z) - C_{qkp} \mathcal{L}[\psi_k(t) \psi_p(t)] = 0 \;,
\end{equation}
where the first term (proportional to $1/z$) vanishes by definition.
Again, considering a further expansion of 
$\psi_q(z) = \psi_q^{(1)} + \psi_q^{(2)}$ with $\psi_q^{(2)}/\psi_q^{(1)} \to 0$ 
as $z \to 0$, we plug into our equation and match orders. 
To the order of $\psi_q^{(1)}$, we obtain
\[
\psi_q^{(1)}(z) = C_{qk} \psi_k^{(1)}(z) \; .
\]
Therefore, we may write $\psi_q^{(1)}(z) = A e_q \phi_\nu(z)$. Plugging this
back into \eqref{MCT4} and multiplying by the left eigenvector of $C_{qk}$ to
eliminate $\psi_q^{(2)}$, we obtain an equation for $\phi_\nu(z)$:
\[
z \phi_\nu^2(z) = \lambda \mathcal{L} [ \phi_\nu^2(t)] \; .
\]

\subsubsection{Glass Side}

Next let us consider what happens when we are slightly away from the critical
density on the glass side. Here we use the ansatz 
$\phi(q, z) = f_q^c/z + (1 - f_q^c)^2 g \psi_q(z)$, with 
$\psi_q(t) \to e_q$ as $t \to \infty$. ($g$ was computed in Section \ref{ENE-Transition}.) 
Starting from \eqref{MCT} and expanding to second order in $g$,
\[
\frac{1}{z}\left( \frac{f_q}{1 - f_q} - F_q[f_k] - C_q\frac{\delta \rho}{\rho} \right)
+ g \left( \psi_q(z) -  C_{qk} \psi_k(z) - \delta C_{qk} \psi_k(z)\right) 
+ g^2\left( z(1 - f_q)\psi_q^2(z) 
- C_{qkp} \mathcal{L}[\psi_k(t) \psi_p(t)] \right) 
= 0 \;.
\]
According to our assumptions the first two terms in the $1/z$ factor cancel.
We are left with
\[
-\frac{1}{z}\left(C_q\frac{\delta \rho}{\rho} \right)
+ g \left( \psi_q(z) -  C_{qk} \psi_k(z) - \delta C_{qk} \psi_k(z) \right) 
+ g^2\left( z(1 - f_q)\psi_q^2(z) 
- C_{qkp} \mathcal{L}[\psi_k(t) \psi_p(t)] \right) 
= 0 \;.
\]
Let us re-write $\delta \rho / \rho$ as $g^2 B$. ($B$ can be determined from the
expression for $g^2$ in Section 4. Note that $\epsilon = \delta \rho / \rho$. Also
note that $\delta C_{qk} = g D_{qk}$ since $\delta C_{qk}$ scales as
$\sqrt{\epsilon}$.) We obtain
\[
  g \left( \psi_q(z) -  C_{qk} \psi_k(z) \right) 
+ g^2\left( z(1 - f_q)\psi_q^2(z) - C_{qkp} \mathcal{L}[\psi_k(t) \psi_p(t)] 
            - D_{qk} \psi_k(z) - \frac{C_q B}{z} \right) 
= 0 \;.
\]
Using the same method as before, we split 
$\psi_k = \psi_k^{(1)} + g \psi_k^{(2)}$. We find
\[
\psi_k^{(1)}(z) = A e_k \psi_\nu(z) \, .
\]
Using this result and eliminating $\psi^{(2)}$ by multiplying by the left
eigenvector we obtain an equation for $\psi_\nu(z)$:
\begin{equation}\label{MCT-Glass}
 z\psi_\nu^2(z) - \lambda \mathcal{L}[\psi_\nu^2(t)] 
            - D' \psi_\nu(z) - \frac{B'}{z} 
= 0 \;,
\end{equation}
which is slightly different from \eqref{MCT-Lambda}.
Now change frequency variables to $\zeta = z/\omega_c$, with $\epsilon \to 0$,
$\omega_c/\epsilon \to 0$ $z \to 0$ and $\zeta = z/\omega_c$ constant. This
implies $\mathcal{L}[f(t)](z) \to \omega_c^{-1} \mathcal{L}[f(\tau)](\zeta)$, 
where $\tau = \omega_c t$. Rewriting according to this change of variables:
\begin{equation}\label{MCT-Glass-Zeta}
 \frac{1}{\omega_c}
  \left\{ \zeta \psi_\nu^2(\zeta) - \lambda \mathcal{L}[\psi_\nu^2(\tau)](\zeta) 
            - D' \psi_\nu(\zeta) - \frac{B'}{\zeta} \right\}
= 0 \;
\end{equation}
We see that we obtain the same equation as before, but with $z \to \zeta$.
Now let us make the ansatz $\psi_\nu(\tau) = 1 + \tau^{-a}$. This implies 
$\psi_\nu(\zeta) = 1/\zeta + \Gamma(1 - a)/\zeta^{1-a}$. Substituting this in 
\eqref{MCT-Glass-Zeta},
\[
 \zeta \left( \frac{\Gamma(1 - a)^2}{\zeta^{2-2a}} + 
               2 \frac{\Gamma(1 - a)}{\zeta^{2 - a}} + \frac{1}{\zeta^2}\right) 
     - \lambda \left( \frac{\Gamma(1 - 2a)^2}{\zeta^{1-2a}} 
           + 2 \frac{\Gamma(1 - a)}{\zeta^{1 - a}} + \frac{1}{\zeta}\right) 
        - D' \left( \frac{\Gamma(1 - a)}{\zeta^{1-a}} + \frac{1}{\zeta} \right) 
                 - \frac{B'}{\zeta} 
= 0 \;.
\]
In the limit $\zeta \gg 1$, $\tau \ll 1$, we find that
\[
\lambda = \frac{\Gamma(1 - a)^2}{\Gamma(1 - 2a)} \; ,
\]
provided $0 < a < 1/2$. In the limit $\zeta \ll 1$ the solution goes to its
final asymptotic value. Plugging in above allows us to obtain an equation 
for $A$ in $\psi_k^{(1)}(t) = A e_k \phi_\nu(t)$, removing some ambiguity from 
our solution. 
Evaluating the limit as $\epsilon \to 0$ and matching the solution with what 
we found at the critical density implies that
\[
\lim_{\epsilon \to 0} \, g (1 - f_k)^2 e_k (1 + \tau^{-a}) 
                                            = (1 - f_k)^2 A e_k t^{-a} \, .
\]
Since $g \sim \epsilon^{1/2}$, we must have
\[
\epsilon^{1/2}\tau^{-a} = \text{constant} \times t^{-a} \,.
\]
From the definition $\tau = \omega_c t$,
\[
\epsilon^{1/2} \omega_c^{-a} = \text{constant} \, .
\]
From this we find that $\omega_c$ scales as $\epsilon^{1/2a}$.

\subsubsection{Liquid Side}

Finally we consider what happens when we are slightly below the critical
density (liquid side). Here we need to assume that a slight change from the
critical density results in a long plateau region which is close to $f_q^c$. 
Following the same procedure as before we obtain a
slightly different equation:
\[
 \zeta \psi_\nu^2(\zeta) - \lambda \mathcal{L}[\psi_\nu^2(t)](\zeta) 
            + D' \psi_\nu(\zeta) + \frac{B'}{\zeta} 
= 0 \;.
\]
Note the change in sign of the final two terms. This is due to the separation 
parameter being negative on the liquid side of the transition,
implying $g$ is negative there as well. However, the
analysis for $\zeta \gg 1$ is basically the same as on the glass side, 
and we obtain
\[
\lambda = \frac{\Gamma(1 - a)^2}{\Gamma(1 - 2a)} \;.
\]
This solution is valid as long as $t_0 \ll t \ll \omega_c^{-1}$, where $t_0$ is
a timescale governing the microscopic dynamics and $\omega_c$ was found in
the previous section to be $\Omega_0 \times \epsilon^{1/2a}$ with $\Omega_0$
constant.

For $\zeta \ll 1$ we consider the ansatz $\psi_\nu(\tau) = 1 -\tau^b$, which
implies $\psi_\nu(\zeta) = 1/\zeta - \Gamma(1 + b) / \zeta^{1 + b}$. Plugging this
into our equation,
\[
 \zeta \left(1 - 2 \frac{\Gamma(1 + b)}{\zeta^{1 + b}} 
                    + \frac{\Gamma(1 + b)^2}{\zeta^{2 + 2b}} \right) 
      - \lambda \left( 1 - 2 \frac{\Gamma(1 + b)}{\zeta^{1 + b}} 
                             + \frac{\Gamma(1 + 2b)}{\zeta^{1 + 2b}} \right) 
            + D'\left( 1 -  \frac{\Gamma(1 + b)}{\zeta^{1 + b}} \right)
                 + \frac{B'}{\zeta} 
= 0 \;.
\]
Taking the limit, and considering only the most singular terms, we obtain
\[
\lambda = \frac{\Gamma(1 + b)^2}{\Gamma(1 + 2b)} \;.
\]
provided $b > 0$.  The validity of our solution is based on the assumption that 
$|z \psi(q, z)| \ll 1$. We used this assumption when we expanded the fraction 
in \eqref{MCT} as a geometric series. Since our solution is singular as $z \to 0$, 
this inequality will eventually be violated rendering our solution invalid. This 
will happen when
\[
g \, \zeta \frac{1}{\zeta^{1 + b}} = 
          g \zeta^{-b} \sim 1 
               \implies \epsilon^{1/2} \omega_c^b z^{-b} \sim 1
                  \implies z \sim \epsilon^{\frac{1}{2b} + \frac{1}{2a}}
                                     = \epsilon^\gamma \, ,
\]
where $\gamma = 1/2a + 1/2b$. Therefore this von Schweidler part of the 
solution is valid when
\[
 \epsilon^{-1/2a} \ll \Omega_0 \times t \ll \epsilon^{-\gamma} \, .
\]
Thus we have analytically found the famous two step process of a power law
to a plateau followed by the von Schweidler decay.
 
\section{Application to FTSPD}

It turns out that the asymptotic analysis of the 
FTSPD is remarkably similar to the MCT analysis. The FTSPD yields
the following kinetic equation for the evolution of the normalized
density autocorrelation function in the case of Smoluchowski Dynamics:
\[
\frac{\partial \phi}{\partial t} + \frac{\bar{D} q^2}{S(q)} \phi(q,t) 
   + \bar{D} q^2 \int_0^t ds \, \rho \beta^2 \Sigma_{BB}(q, t - s) \frac{\partial \phi(q, s)}{\partial s} = 0
\]
where the memory function kernel is given at second order by
\begin{subequations}
\begin{equation}\label{SigmaBB}
\rho^4 \beta^2 \Sigma_{BB}(q, t) = \frac{1}{2} \int \frac{d^3 k}{(2 \pi)^3}
    \left[ \tilde{G}_{\rho \rho}(k, t)G_{\rho \rho}^{(0)}(|\vec{q} - \vec{k}|, t) + \right. 
        \left. \bar{G}_{\rho \rho}(k, t) \bar{G}_{\rho \rho}(|\vec{q} - \vec{k}|, t) \right] \, .
\end{equation}
In writing down \eqref{SigmaBB} we have already assumed that the frequency
dependence of the 3-point vertices can, reasonably, be dropped in the 
long-time limit. This amounts to dropping time derivative terms
acting on the density-related quantities $G^{(0)}_{\rho \rho}$, $G_{\rho \rho}$,
$\tilde{G}_{\rho \rho}$, $\bar{G}_{\rho \rho}$ where $G^{(0)}_{\rho \rho}$
is the non-interacting correlation function, and the dressed propagators
$\tilde{G}_{\rho \rho}$ and $\bar{G}_{\rho \rho}$ are discussed in some
detail in Appendix A. In the following section we demonstrate 
that this frequency dependence is irrelevant for our asymptotic analysis. 
To analyze the asymptotic behavior of this equation using the 
methods of MCT we must examine the Laplace transform
of $\phi(q, t)$:
\[
z\phi(q, z) - 1 + \frac{\bar{D} q^2}{S(q)} \phi(q, z) 
      + \bar{D} q^2 \mathcal{L}[\rbs \Sigma_{BB}(q, t)](z\phi(q, z) - 1) = 0 \, ,
\]
where we have used the initial condition $\phi(q, 0) = 1$ and 
$\mathcal{L}[\Sigma_{BB}(q, t)]$ is the Laplace transform of $\Sigma_{BB}(q, t)$. 
Dividing by $z \phi(q, z) - 1$ we obtain
\begin{equation}\label{FTSPD-Asymptotic}
\frac{\bar{D} q^2}{S(q)} \frac{\phi(q, z)}{1 - z \phi(q, z)} 
        = 1 + \bar{D} q^2 \mathcal{L}[ \rbs \Sigma_{BB}(q, t)] \,.
\end{equation}
\end{subequations}
We see that we obtain essentially the same equation as in MCT but with a different kernel.

\subsection{Long-time behavior of $\Sigma_{BB}(q, t)$}

For the asymptotic analysis of our kinetic equation \eqref{FTSPD-Asymptotic}, we must determine which terms
in the kernel $\Sigma_{BB}(q,t)$ are relevant at long times. We begin the analysis
with an examination of $\tilde{G}_{\rho \rho}$, defined in the frequency domain by
\begin{align*}
\tilde{G}_{\rho \rho}(q, \omega) 
         &= G^{(0)}_{\rho i} \sigma_{ij} G_{jk} \sigma_{k \ell} G^{(0)}_{\ell \rho} \\
         &= V(q)^2 \left\{ G^{(0)}_{\rho \rho} G_{B \rho} G^{(0)}_{B \rho} + G^{(0)}_{\rho B} G_{\rho \rho} G^{(0)}_{B \rho} + G^{(0)}_{\rho B} G_{\rho B} G^{(0)}_{\rho \rho}\right\} \; ,
\end{align*}
where $\sigma_{ij} = V(k)$ if $i \neq j$ and is zero otherwise \cite{SDENE}. 
In the time domain this becomes
\begin{multline*}
\tilde{G}_{\rho \rho}(q, t)
        = V(q)^2 \int_{-\infty}^\infty ds  \int_{-\infty}^\infty ds'
                       \left\{ \GrrO(t - s' - s) \Gbr(s) \GbrO(s')   
                      +  \GrbO(t - s' - s) \Grr(s) \GbrO(s') \right.\\
                      + \left. \GrbO(t - s' - s) \Grb(s) \GrrO(s') \right\} ,
\end{multline*}
where we have suppressed the wave-number dependence on the right-hand side. We shall
now examine each of the three terms in the integrand in turn. Using the fluctuation
dissipation theorem, we have for the first term
\begin{align*}
\beta^{-2} \int_{-\infty}^\infty ds  \int_{-\infty}^\infty ds' \, \GrrO(t - s' - s) \Gbr(s) \GbrO(s') &=
\int_{-\infty}^\infty ds  \int_{-\infty}^\infty ds' \, 
          \GrrO(t - s' - s) \theta(-s) \dot{G}_{\rho\rho}(-s) \theta(-s') \dot{G}_{\rho\rho}^{(0)}(-s') \\
&= \int_{-\infty}^0 ds  \int_{-\infty}^0 ds' \, 
          \GrrO(t - s' - s) \dot{G}_{\rho\rho}(-s) \dot{G}_{\rho\rho}^{(0)}(-s') \\
&= \frac{1}{2} \GrrO(0) \int_{-\infty}^0 ds \, 
          \GrrO(t - s) \dot{G}_{\rho\rho}(s) \, ,
\end{align*}
where we have performed the integral over $s'$ in the final line.
For the second term we have
\begin{multline*}
\beta^{-2} \int_{-\infty}^\infty ds  \int_{-\infty}^\infty ds' \, \GrbO(t - s' - s) \Grr(s) \GbrO(s') \\
\begin{aligned}
&= \int_{-\infty}^\infty ds  \int_{-\infty}^\infty ds' \, 
 \theta(t - s' - s) \dot{G}_{\rho\rho}^{(0)}(t - s' - s) \Grr(s) \theta(-s') \dot{G}_{\rho\rho}^{(0)}(-s') \\
&= \int_{-\infty}^0 ds'  \int_{-\infty}^{t - s'} ds \, 
 \dot{G}_{\rho\rho}^{(0)}(t - s' - s) \Grr(s) \dot{G}_{\rho\rho}^{(0)}(-s') \\
&= \int_{-\infty}^0 ds'  \int_{-\infty}^{t - s'} ds \, 
 \GrrO(t - s' - s) \dot{G}_{\rho\rho}(s) \dot{G}_{\rho\rho}^{(0)}(-s') 
         - \int_{-\infty}^0 ds' \, \GrrO(0) \Grr(t - s') \dot{G}_{\rho\rho}^{(0)}(-s') \\
&= - \bar{D} q^2 \int_{-\infty}^0 ds'  \int_{-\infty}^{t - s'} ds \, 
 \GrrO(t - s' - s) \dot{G}_{\rho\rho}(s) \GrrO(s') 
         + \GrrO(0) \int_{t}^\infty \GrrO(t - s) \dot{G}_{\rho\rho}(s) \, ds
\end{aligned}\\
+ \GrrO(0)^2 \Grr(t) \, .
\end{multline*}
Finally, the third term becomes
\begin{align*}
\beta^{-2} \int_{-\infty}^\infty ds  \int_{-\infty}^\infty ds' \, \GrbO(t - s' - s) \Grb(s) \GrrO(s') &=
\int_{-\infty}^\infty ds  \int_{-\infty}^\infty ds' \, 
          \theta(t - s' - s) \dot{G}_{\rho\rho}^{(0)}(t - s' - s) \theta(s) \dot{G}_{\rho\rho}(s) \GrrO(s') \\
&= \int_0^\infty ds  \int_{-\infty}^{t - s} ds' \, 
          \dot{G}_{\rho\rho}^{(0)}(t - s' - s) \dot{G}_{\rho\rho}(s) \GrrO(s') \\
&=  -\bar{D} q^2 \int_0^\infty ds  \int_{-\infty}^{t - s} ds' \, 
          \GrrO(t - s' - s) \dot{G}_{\rho\rho}(s) \GrrO(s') \, .
\end{align*}
Next, using the fact that $\dot{G}_{\rho\rho}^{(0)}(t) = \pm \bar{D} q^2 \GrrO(t)$ to add up 
the double integrals from the second and third terms (with the sign in front of $q^2$ 
depending on the sign of $t$), we obtain
\begin{multline*}
- \bar{D} q^2 \int_{-\infty}^0 \!\!\!\! ds'  \int_{-\infty}^{t - s'} \!\!\!\! ds \, 
 \GrrO(t - s' - s) \dot{G}_{\rho\rho}(s) \GrrO(s')
- \bar{D} q^2 \int_0^\infty \!\!\!\! ds  \int_{-\infty}^{t - s} \!\!\!\! ds' \, 
          \GrrO(t - s' - s) \dot{G}_{\rho\rho}(s) \GrrO(s')
\\
= - \bar{D} q^2 \int_0^t ds \int_0^{t - s} ds' \GrrO(t - s - s') \dot{G}_{\rho \rho}(s) \GrrO(s') 
  - \GrrO(0) \int_0^\infty \GrrO(t - s) \dot{G}_{\rho \rho}(s) \, ds \\ 
  - \frac{1}{2} \GrrO(0) \int_{-\infty}^0 ds \, 
          \GrrO(t - s) \dot{G}_{\rho\rho}(s) \, .  
\end{multline*}
Adding all three terms,
\begin{align*}
\tilde{G}_{\rho \rho}(q, t) &= [\beta V(q)]^2 \left\{ 
                       - \bar{D} q^2 \int_0^t ds \int_0^{t - s} ds' \GrrO(t - s - s') \dot{G}_{\rho \rho}(s) \GrrO(s') \right.\\
                            &\hspace{10em} \left. - \; \GrrO(0) \int_0^t \GrrO(t - s) \dot{G}_{\rho \rho}(s) \, ds 
                           \; + \; \GrrO(0)^2 \Grr(t) \right\}\\
&= [\beta V(q)]^2 \GrrO(0) \left\{ \GrrO(0) \Grr(t) - 
                      \int_0^t ds \, [1 + \bar{D} q^2(t - s)] \GrrO(t - s) \dot{G}_{\rho \rho}(s) \right\} \, .
\end{align*}
From here we compute the Laplace transform,
\begin{align*}
\tilde{G}_{\rho \rho}(q, z) &= [\beta V(q)]^2 \GrrO(t = 0) \left\{
        \GrrO(t = 0) \Grr(z) - \frac{z + 2 \bar{D} q^2}{(z + \bar{D} q^2)^2} (z\Grr(z) - \Grr(t = 0))\right\} \, .
\end{align*}
We see that the only singularities as $z \to 0$ come from $\Grr(z)$. Therefore, in the $z \to 0$ limit
we find that the most dominant term is
\[
\tilde{G}_{\rho \rho}(q, z) \to [\beta V(q) \rho_0]^2 \Grr(q, z) \, ,
\]
where we have replaced $\GrrO(0)$ with $\rho_0$. Taking the inverse transform,
\[
\tilde{G}_{\rho \rho}(q, t \gg t_0) = [\beta V(q) \rho_0]^2 \Grr(q, t) \, ,
\]
where $t_0$ is a timescale governing the early-time dynamics. Plugging in our ansatz, 
\mbox{$G_{\rho \rho}(q, t) = S(q)(f_q + h_q|t|^\alpha)$}, and multiplying by $\GrrO(t)$,
\begin{align*}
\mathcal{L}[ \tilde{G}_{\rho \rho}(q, t)\GrrO(q, t)](z) 
         &= [\beta V(q) \rho_0]^2 S(q) \mathcal{L}[ (f_q + h_q t^\alpha) \GrrO(t) ] \\
         &= [\beta V(q) \rho_0]^2 S(q) \rho_0 \left\{f_q\frac{1}{z + \bar{D} q^2} 
                            + h_q \frac{\Gamma(1 + \alpha)}{(z + \bar{D} q^2)^{1 + \alpha}} \right\} \, .
\end{align*}
Since this term is not singular as $z \to 0$ it does not affect 
the asymptotic analysis in the lowest order limit. Thus we can ignore 
the ``single particle'' contribution to $\Sigma_{BB}$ in \eqref{SigmaBB}. 
Note that the derivative terms in the kernel add multiples of $z$ to 
our Laplace transformed expressions which implies that they go to zero in the limit. It follows that 
the dominant contribution to the asymptotic equation from the kernel is the ``collective''
$\bar{G}_{\rho \rho}(q - k, t)\bar{G}_{\rho \rho}(k, t)$ term.

\subsection{Ergodic/Non-ergodic Transition}
To determine if an ENE transition occurs we set $\phi(q, z) = f_q/z$ and let $z \to 0$.
Plugging into \eqref{FTSPD-Asymptotic}, we obtain
\[
\frac{\bar{D} q^2}{S(q)} \frac{f_q/z}{1 - f_q} = 1 + \bar{D} q^2 \mathcal{L}[\rbs \Sigma_{BB}(q, t)]
\]
with
\[
\mathcal{L}[\rbs \Sigma_{BB}] 
       = \frac{1}{z} \cdot \frac{\beta^2}{2 \rho} 
           \int \frac{d^3 k}{(2 \pi)^3} V(k)S(k) V(q - k)S(q - k) f_k f_{q - k} \, .
\]
Putting this form into the equation, and setting
\[
F_q[f_k] = S(q)\frac{\beta^2}{2 \rho} \int \frac{d^3 k}{(2 \pi)^3} V(k)S(k)V(q - k)S(q - k) f_k f_{q - k}
\]
we obtain
\[
\frac{1}{z} \frac{f_q}{1 - f_q} = \frac{S(q)}{\bar{D} q^2} + \frac{1}{z} F_q[f_k]
\]
which becomes
\[
\frac{f_q}{1 - f_q} = F_q[f_k]
\]
in the $z \to 0$ limit. From here it is apparent that the determination of the 
ENE transition is identical to MCT, except that we have a different ``vertex''
or potential.

\subsection{Time Dependence at Transition}
Ignoring the constant term (which becomes negligible in the $z \to 0$ limit), equation 
\eqref{FTSPD-Asymptotic} reduces to
\[
\frac{\phi(q, z)}{1 - z \phi(q, z)} = \mathcal{L}[\rbs \Sigma_{BB}] \,.
\]
We apply the ansatz $\phi(q, t) = f(q) + (1 - f(q))^2 \psi_q(t)$. The 
Laplace transform of the ansatz is given by
\[
\phi(q, z) = \frac{f(q)}{z} + (1 - f(q))^2\psi_q(z) \; .
\]
We now plug this into the equation and expand the left hand side to second order in $\psi_q(z)$:
\[
\frac{1}{1 - f_q} \left( \frac{f_q}{z} + (1 - f_q)\psi_q(z) + z 
(1 - f_q)^2 \psi_q^2(z) \right) = \mathcal{L}[\rbs \Sigma_{BB}(q, t)] \; .
\]
To apply the ansatz to the right hand side we must analyze $\bar{G}_{\rho \rho}$.
Ultimately, we need to compute the Laplace transform of 
$\bar{G}_{\rho \rho}(q,t)\bar{G}_{\rho \rho}(p, t)$. A
prerequisite for this computation is $\bar{G}_{\rho \rho}(q, t)$. Our ansatz will be 
\[
G_{\rho \rho}(q, t) = S(q) \left( f_q + h_q |t|^\alpha \right) \, .
\]
This implies we set $\psi_q(t) = h_q |t|^\alpha / (1 - f_q)^2 $. 
\begin{align*}
\bar{G}_{\rho \rho}(q, \omega)&= \frac{V(q)}{2} \left[ (G^{(0)}_{\rho \rho}(q, \omega) ( G_{\rho B}(q, \omega) 
               + G_{B \rho}(q, \omega))  
               + G_{\rho \rho}(q, \omega) ( G_{\rho B}^{(0)}(q, \omega) + G_{B \rho}^{(0)}(q, \omega)) \right] \, .
\end{align*}
In the time domain, we have (suppressing wave-number arguments)
\begin{align}
\bar{G}_{\rho \rho}(t) &= \frac{V(q)}{2} \int_{-\infty}^\infty \, ds \left[ (\GrrO(s) ( \Grb(t - s) + \Gbr(t - s))
               + \Grr(s) ( \GrbO(t - s) + \GbrO(t - s)) \right] \notag\\ 
               &= \frac{\beta V(q)}{2} \left\{ \int_t^\infty  \dot{G}_{\rho \rho}(s - t) \GrrO (s) \; ds 
                      + \int_{-\infty}^t  \dot{G}_{\rho \rho}(t - s) \GrrO (s) \; ds  \right. \notag\\
& \hspace{20em} +\left. \beta^{-1} \int_{-\infty}^\infty  
                    G_{\rho \rho}(t - s) ( G_{\rho B}^{(0)}(s) + G_{B \rho}^{(0)}(s)) \; ds \right\}
                      \label{GrhoTime}
\end{align}
where we have invoked the fluctuation dissipation theorem (FDT) in the second line.
Using the FDT on one of the terms in the third integral we obtain
\begin{align*}
\beta^{-1}\int_{-\infty}^\infty G_{\rho \rho}(t - s) G_{B \rho}^{(0)}(s) \; ds 
                  &= \int_{-\infty}^0 \Grr(t - s) \dot{G}^{(0)}_{\rho \rho}(-s) \; ds \\
                  &= -\Grr(t)\GrrO(0) - \int_{-\infty}^0 \dot{G}_{\rho \rho}(t - s) \GrrO(-s) \; ds \\
                  &= -\Grr(t)\GrrO(0) - \int_{-\infty}^0 \dot{G}_{\rho \rho}(t - s) \GrrO(s) \; ds \, ,
\end{align*}
where we have used the properties of $\GrrO(t)$ in the second and third lines.
A similar computation reveals that the other term in the third integral of 
\eqref{GrhoTime} reduces to
\begin{align*}
\beta^{-1} \int_{-\infty}^\infty \Grr(t - s) \GrbO(s) \; ds  = 
          -\Grr(t)\GrrO(0) + \int_0^\infty \dot{G}_{\rho \rho}(t - s) \GrrO(s) \; ds \, .
\end{align*}
Adding all the pieces together,
\begin{align*}
\bar{G}_{\rho \rho}(q, t) &= \frac{\beta V(q)}{2} \left\{ 
                        \int_t^\infty  \dot{G}_{\rho \rho}(s - t) \GrrO (s) \; ds 
                      + \left[ \int_{-\infty}^t + \int_0^\infty - \int_{-\infty}^0 \right] 
                            \dot{G}_{\rho \rho}(t - s) \GrrO(s) \; ds
                      - 2 \Grr(t)\Grr(0) \right\}\\
                 &=  \beta V(q) \left\{ - \Grr(t)\GrrO(0) 
                                        + \int_0^t \dot{G}_{\rho \rho}(t - s) \GrrO(s) \; ds \right\} \, .
\end{align*}
Taking the Laplace transform,
\begin{align*}
\bar{G}_{\rho \rho}(q, z) = \beta V(q) \left\{ - \Grr(z)\GrrO(t = 0) 
                                        + (z \Grr(z) - \Grr(t = 0)) \GrrO(z) \right\} \, .
\end{align*}
Since $\GrrO(z)$ is not singular as $z \to 0$, we find that the non-integral term of
$\bar{G}_{\rho \rho}(t)$ dominates in the long-time limit. 
Therefore, we may replace
\begin{align*}
\bar{G}_{\rho \rho}(q, t) \to -\beta \rho_0 V(q) S(q) (f_q + h_q |t|^\alpha)
\end{align*}
in this limit.
At this point the analysis becomes identical to the MCT case, except that
we have a different vertex function. Using the notation of Section 3, let
\begin{align*}
F_q[f_k] &=  \frac{\beta^2 S(q)}{2 \rho}  \int \frac{d^3 k}{(2 \pi)^3} 
V(k) S(k) V(q - k) S(q - k) f_k f_{q - k} \\
C_{q} &= \rho \frac{\delta F_q[f_k]}{\delta \rho}\\
C_{qk} &= \frac{\delta F_q[f_k]}{\delta f_k} (1 - f_k)^2 \\
C_{qkp} &= \frac{1}{2}\frac{\delta^2 F_q}{\delta f_k \delta f_p} 
(1 - f_k)^2 (1 - f_p)^2
\end{align*}
where $V(q)$ is a pseudo-potential determined in \cite{SDENE}.
Putting all this together, we find
\[
\lim_{z \to 0} \mathcal{L}[\Sigma_{BB}(q,t)](z) = \frac{F_q[f_k]}{z} + C_{qk}\psi_k(z) 
              + C_{qkp} \mathcal{L}[\psi_k(t) \psi_p(t)] \, .
\]
Therefore,
\[
\frac{1}{z}\left( \frac{f_q}{1 - f_q} -  F_q[f_k] \right) + \left( \psi_q(z) -  C_{qk} \psi_k(z) \right) + z(1 - f_q)\psi_q^2(z) - C_{qkp} \mathcal{L}[\psi_k(t) \psi_p(t)] = 0 \;,
\]
and the analysis for the ergodic and non-ergodic regimes now proceeds identically to the mode 
coupling case, the only difference being the ``vertex'' in MCT and the pseudo-potential in
FTSPD. 
\begin{figure}[h]
  \centering
  \input{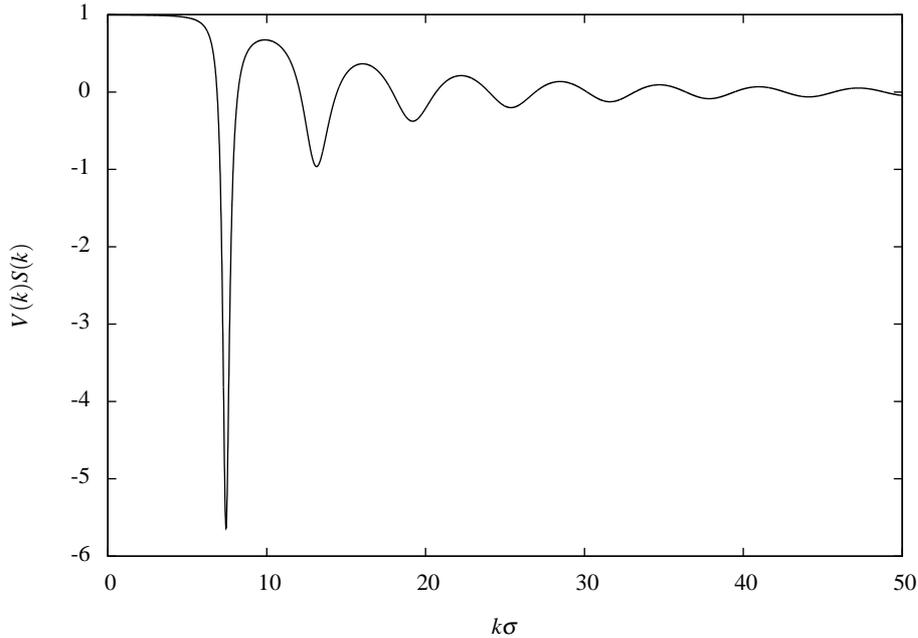}
  \caption{The product $V(k)S(k)$ appearing in the FTSPD kinetic equation kernel 
    $\Sigma_{BB}(q,t) = \frac{\pi}{12 \eta} \int \frac{d^3 k}{(2 \pi)^3}
    V_k S_k V_{q - k} S_{q - k} 
    \bar{\phi}(k, t) \bar{\phi}(q - k, t)$ at the transition.
    The negative values of this combination imply that $C_{qk}^{\text{FT}}$ is not a 
    positive matrix. This is why we cannot choose $\hat{e}_k$ such that $\hat{e}_k > 0$
    for all $k$ as we can in Mode Coupling Theory. Nevertheless, the maximum eigenvalue
    of $C_{qk}^{\text{FT}}$ approaches unity as the packing fraction tends to the transition 
    point implying the analysis is still valid.
  }
  \label{fig:pseudo}
\end{figure}
\begin{figure}[h]
  \centering
  \input{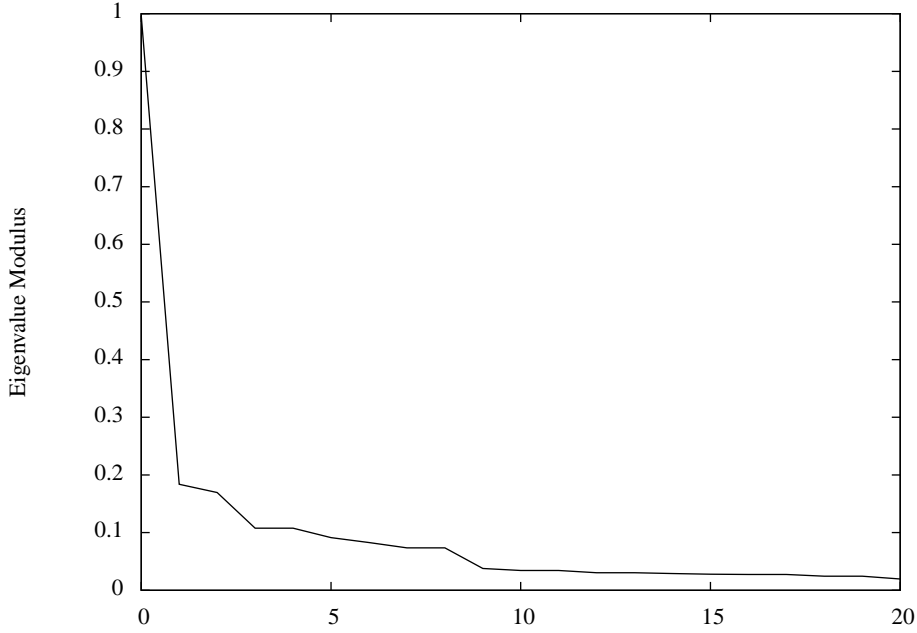}
  \caption{The eigenvalues of $C_{qk}^\text{FT}$ sorted by eigenvalue modulus. The largest
    eigenvalue is real, non-degenerate, and approaches unity as the density approaches the 
    transition point.
    This graph was constructed from $C_{qk}^\text{FT}$ computed on a $800 \times 800$ grid
    with a spacing $dk = 0.1$. The spectra for grids down to $400 \times 400$ with $dk = 0.2$
    and up to $1600 \times 1600$ with $dk = 0.05$ are essentially identical. 
  }
  \label{fig:eigen}
\end{figure}

\section{Numerical Results for Hard Spheres}


Solving for $f$ and $\psi$ as suggested by the preceding MCT analysis yields results
in close agreement with \cite{BGL}. We find that with the Percus-Yevick
approximation for the structure factor of hard spheres the transition happens 
at a packing fraction of $\eta_{\text{MCT}} = 0.515$ and $\lambda_{\text{MCT}} = 0.734$.
Solving for the exponents, we find $a = 0.312$ and $b = 0.584$. 
We find that $\lambda$ increases quite sharply in the vicinity of the critical density.
Barrat et. al. report $\eta_{\text{MCT}} = 0.525$ and $\lambda_{\text{MCT}} = 0.758$ with 
the Verlet-Weiss approximation.

While conducting these numerical experiments it became clear that convergence
near the critical density requires integrating out to fairly high 
($k\sigma \sim 100$) wavenumbers. Increasing the range beyond this does not have
a noticeable effect on the calculations. In addition, the number of iterations 
required for the convergence of the non-ergodicity parameter increases quite 
rapidly as the transition density is approached. For instance, convergence to 
three digits in the eigenvalue occurs after 60 iterations for $\eta_{\text{MCT}} = 0.52$, 
but it takes over 700 iterations at $\eta_{\text{MCT}} = 0.51505$.

\begin{figure}[h]
  \centering
  \input{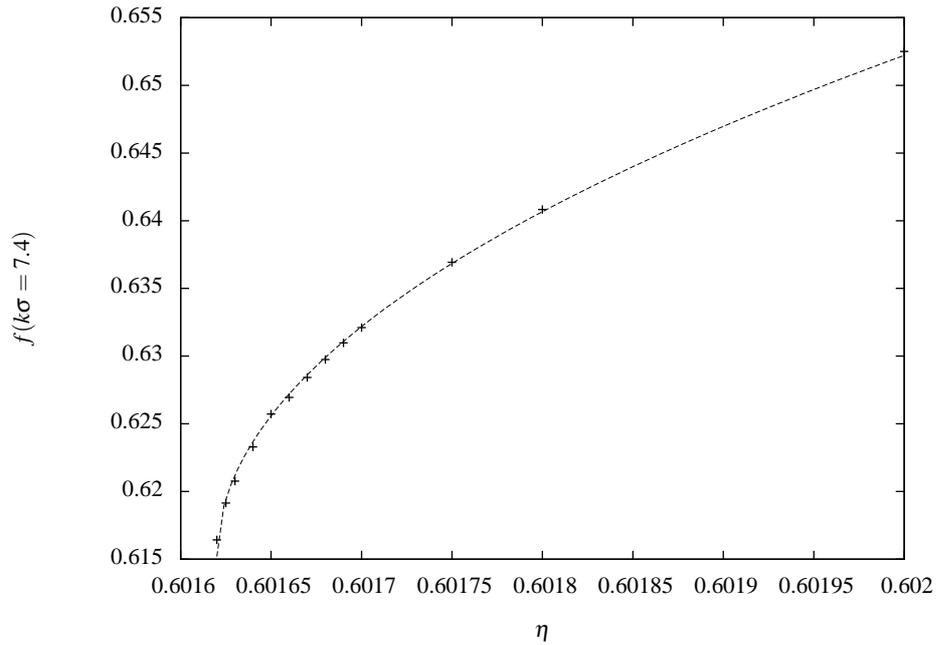}
  \caption{This plot shows $f^{\text{FT}}(k \sigma = 7.4)$, the non-ergodicity parameter at
    wavenumber $k \sigma = 7.4$, as a function of the packing fraction. The 
    dashed line shows a least squares fit of $A\sqrt{\epsilon} +  B$ to the data.}
  \label{fig:f_kscaleft}
\end{figure}

The vertex in MCT has some special properties which allow one to draw
conclusions about the eigenvalues and eigenvectors. Because $C^{MCT}_{qk}$ is a positive matrix
it can be proved that there is a largest eigenvalue and that it becomes unity in the
continuum limit. In addition, it can be shown that the both the right and left eigenvectors, 
$e_k$ and $\hat{e}_k$, must be positive. In FTSPD the situation is more complicated - as can be
seen from Figure \ref{fig:pseudo} the pseudopotential no longer guarantees that $C^{FT}_{qk} \geq 0$. 
Therefore the same proofs from MCT cannot be carried over to the FTSPD case. However, despite the 
complication with the pseudopotential, numerical evidence (see Figure \ref{fig:eigen}) indicates 
that the largest eigenvalue of $C^{FT}_{qk}$ is unity. In addition, the right eigenvector $e_k$ 
is found to be positive, as in MCT. We also 
observe the same $\sqrt{\epsilon}$ scaling behavior in $f_q^{FT}$ that we see in MCT 
(see Figure \ref{fig:f_kscaleft}).

\begin{figure}[h]
  \centering
  \input{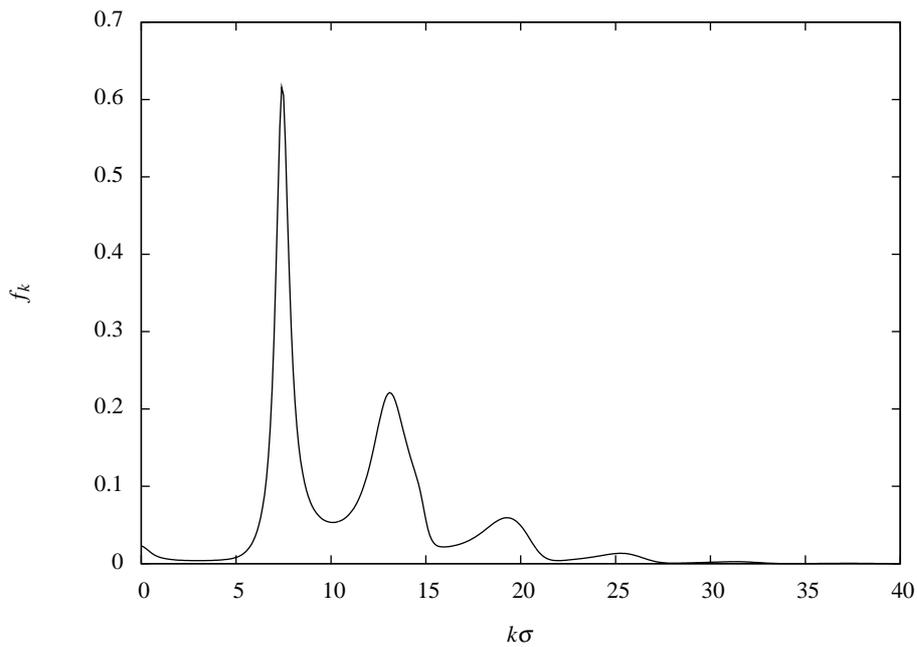}
  \caption{This plot shows $f_k^{\text{FT}}$ at the critical density $\eta_{\text{FT}} = 0.602$.}
  \label{fig:f_kft}
\end{figure}

\begin{figure}[h]
  \centering
  \input{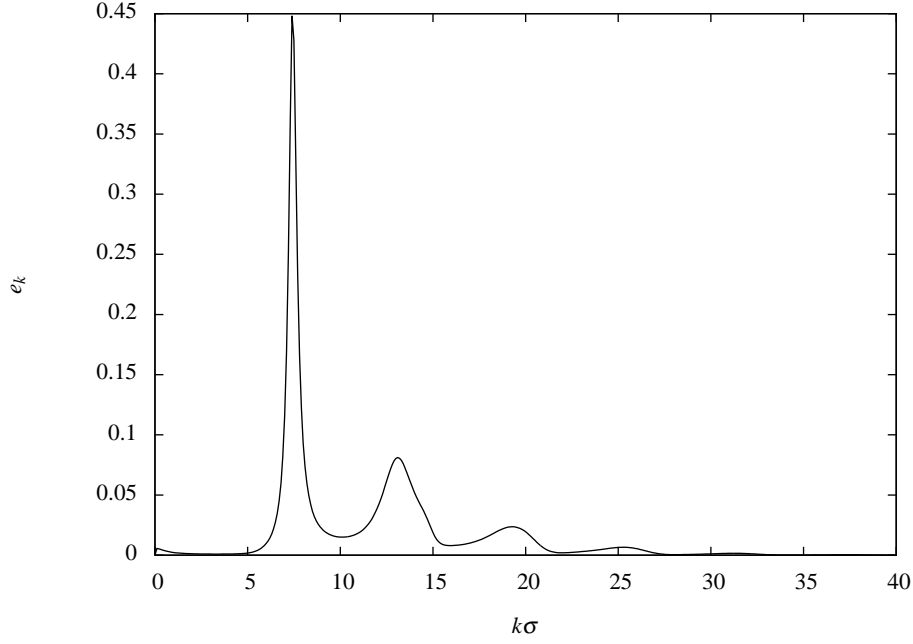}
  \caption{This plot shows $e_k^{\text{FT}}$, the right eigenvector of $C_{qk}^{\text{FT}}$, at 
    the critical density $\eta_{\text{FT}} = 0.602$. The eigenvector is normalized
    according to the convention $\sum_k \hat{e}_k \hat{e}_k = 1$. }
  \label{fig:e_kft}
\end{figure}

\begin{figure}[h]
  \centering
  \input{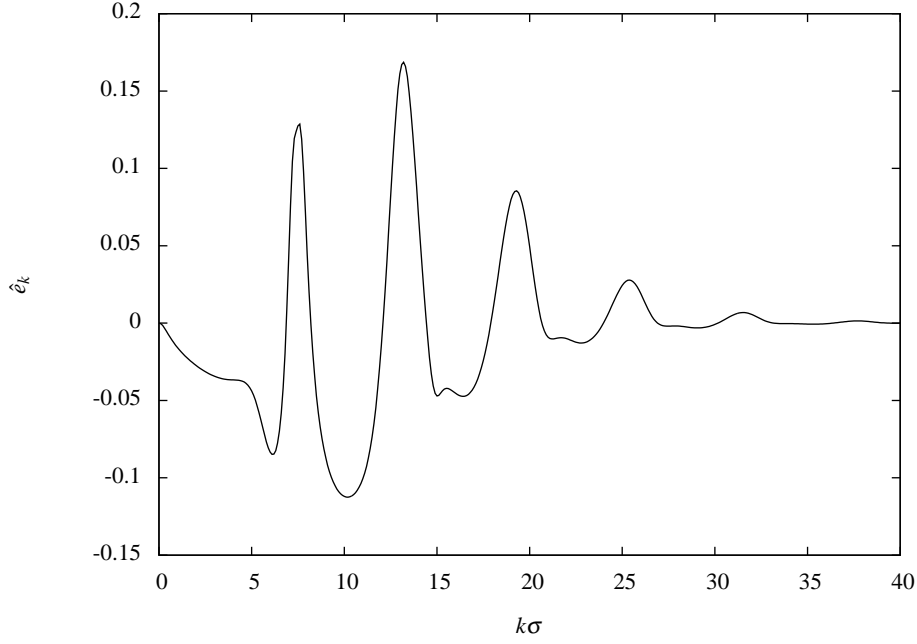}
  \caption{This plot shows $\hat{e}_k^{\text{FT}}$, the left eigenvector of $C_{qk}^{\text{FT}}$, at 
    the critical density $\eta_{\text{FT}} = 0.602$. We see that this eigenvector is not
    positive for all $k$, in contrast to MCT. The reason for this is that $C_{qk}^{\text{FT}}$ is
    not a positive matrix, rendering the Perron-Frobenius theorems inapplicable. The eigenvector is 
    normalized according to the convention  $\sum_k \hat{e}_k \hat{e}_k = 1$. }
  \label{fig:e_khft}
\end{figure}

Using the first-order result
for the pseudopotential for simplicity, which is essentially the direct 
correlation function \cite{FTSPD, SDENE},
solving the FTSPD for hard spheres with the Percus-Yevick approximation
appears to be computationally easier than MCT. Convergence only requires
integration to at most $k \sigma \sim 80$ compared to $k \sigma \sim 100$
for MCT. This is expected because the decay of the vertex function is
much faster for large wavenumbers in the FTSPD. Convergence requires more
iteration closer to the transition similarly to MCT.  We find that the ENE
transition occurs at a packing fraction of $\eta_{\text{FT}} = 0.602$
with a corresponding $\lambda_{\text{FT}} = 0.553$. The non-ergodicity parameter
at the transition is plotted in Figure \ref{fig:f_kft}, while the right and left
eigenvectors are plotted in Figures \ref{fig:e_kft} and \ref{fig:e_khft}, respectively.
Solving for the exponents, we obtain $a = 0.386$ for the power law relaxation 
and $b = 0.935$ for the von Schweidler relaxation.
 
\section{Conclusion}


We have demonstrated that, to second order in the effective potential, the
FTSPD essentially reduces to a mode coupling theory. This shows that FTSPD is a 
framework from which MCT can be derived and systematic corrections introduced. 
We find that an ergodic / non-ergodic transition occurs at a packing fraction of
$\eta_{\text{FT}} = 0.602$, which is in good agreement with some recent
simulations and experiments \cite{BrambillaSim}. This agreement prompts us to investigate
the effectiveness of the pseudopotential approach for other systems. In addition, we plan to 
confirm this asymptotic analysis by a direct integration of the FTSPD kinetic equation. 
Although we have gathered extensive numerical evidence for their veracity, proofs that the 
largest eigenvalue $E_0$ of $C_{qk}^{\text{FT}}$ is real, positive, and nondegenerate and $E_0 \to 1$ as 
$\eta \to \eta^*$ are still unavailable to the best of our knowledge. In a forthcoming
paper we extend this analysis to a treatment of binary mixtures. An analysis of 
Newtonian dynamics within the framework of FTSPD is given in \cite{Newt1}.

\begin{appendices}

\section{Properties of $\tilde{G}$ and $\bar{G}$}

In this appendix we prove that the dressed propagators 
$\bar{G}$ and $\tilde{G}$ individually satisfy the
same fluctuation dissipation theorem that $G$ satisfies:
\[
\Grr(q, \omega) = \frac{2}{\beta \omega} \im \Gbr(q, \omega) \, .
\] 
In the time domain, 
\[
\Gbr(q, t) = \beta \theta(-t) \Grr(q, -t) \, .
\]
This result holds at all orders of perturbation theory. The 
non-interacting cumulants are given by
\[
\GrrO(q, \omega) = \frac{2\rho_0 \bar{D} q^2}{\omega^2 + (\bar{D} q^2)^2} \\
\]
and
\[
\GrbO(q, \omega) = \GbrO(q, \omega)^* = \frac{\beta \rho_0 \bar{D} q^2 }{i \omega - \bar{D}q^2} \, .
\]
In the following subsections it will be useful to
decompose the cumulants into real and imaginary
components:
\begin{align*}
G^{(0)}_{B \rho}    &= R_0 + iI_0             &G_{B \rho} &= R  + iI   \\
G^{(0)}_{\rho B}    &= R_0 - iI_0             &G_{\rho B} &= R  - iI   \\
G^{(0)}_{\rho \rho}  &= \frac{2}{\beta \omega} I_0  &G_{\rho \rho} &= \frac{2}{\beta \omega} I
\end{align*}

\subsection{$\bar{G}$ FDS}

Recall the form of $\bar{G}_{ij}$ given by
\[
\bar{G}_{ij} = 
 \frac{1}{2}
  (G^{(0)}_{ix} \sigma_{xy} G_{yj} + G_{ix} \sigma_{xy} G^{(0)}_{yj} )
\]
Explicitly, this yields
\begin{align*}
\bar{G}_{\rho B} &= 
   \frac{1}{2}(G^{(0)}_{\rho B} \beta V G_{\rho B} + G_{\rho B} \beta V G^{(0)}_{\rho B} ) 
        = G^{(0)}_{\rho B} \beta V G_{\rho B} \, , \\
\bar{G}_{B \rho} &= 
\frac{1}{2}(G^{(0)}_{B \rho} \beta V G_{B \rho} + G_{B \rho} \beta V G^{(0)}_{B \rho} ) 
        = G^{(0)}_{B \rho} \beta V G_{B \rho} \, ,
\end{align*}
and
\[
\bar{G}_{\rho \rho} = 
  \frac{1}{2}(G^{(0)}_{\rho B} \beta V G_{\rho \rho} + 
              G^{(0)}_{\rho \rho} \beta V G_{B \rho} + 
              G_{\rho B} \beta V G^{(0)}_{\rho \rho} + 
              G_{\rho \rho} \beta V G^{(0)}_{B \rho} ) \, .
\]
If we write out our contributing terms as real and imaginary components
then we have for the imaginary part of $\bar{G}_{\rho B}$,
\begin{align*}
\im \bar{G}_{B \rho} &= \frac{1}{2i}(\bar{G}_{B \rho} - \bar{G}_{\rho B})    \\
&= \frac{\beta V}{2i}\left[ (R_0 + iI_0 )(R + iI) - (R_0 - iI_0 )(R - iI) \right] \\
&= \beta V (IR_0 + I_0 R) \, .
\end{align*}
Looking next at $\bar{G}_{\rho \rho}$, we have
\[
\bar{G}_{\rho \rho} = \frac{2 V}{\omega}(R_0 I + R I_0) \, .
\]
Finally, we obtain the fluctuation dissipation relation
\[
\bar{G}_{\rho \rho} = \frac{2}{\beta \omega} \im \, \bar{G}_{B \rho} \, .
\]

\subsection{$\tilde{G}$ FDS}

We may repeat the same procedure for $\tilde{G}_{ij}$, which is given by
\begin{align*}
\tilde{G}_{ij} &= G^{(0)}_{ix} \sigma_{xy} G_{yz} \sigma_{zp} G^{(0)}_{pj} \, .
\end{align*} 
This implies
\[
 \tilde{G}_{\rho B} = \tilde{G}_{B \rho}^* =  \GrbO \beta V \Grb \beta V \GrbO 
\]
and
\[ 
\tilde{G}_{\rho \rho} 
  = \GrrO \beta V \Gbr \beta V \GbrO + \GrbO \beta V \Grr \beta V \GbrO 
             + \GrbO \beta V \Grb \beta V \GrrO \, . 
\]
Using the same decomposition into real and imaginary components, we have
\begin{align*}
\im \, \tilde{G}_{B \rho} 
      &= \frac{1}{2i}\left( \tilde{G}_{B \rho} - \tilde{G}_{\rho B} \right) \\
      &= \frac{1}{2i}\left(\GbrO \beta V \Gbr \beta V \GbrO - \GrbO \beta V \Grb \beta V \GrbO \right)\\
      &= \frac{(\beta V)^2}{2i}\left[ (R_0 + iI_0)(R + iI)(R_0 + iI_0) \right. 
             \left. - (R_0 - iI_0)(R - iI)(R_0 - iI_0) \right] \\
      &= (\beta V)^2 \left( (R_0^2 - I_0^2)I + 2 R_0I_0 R \right)
\end{align*}
and
\begin{align*}
\tilde{G}_{\rho \rho} &= \frac{2\beta V^2}{\omega} 
            \left[ I_0(R + iI)(R_0 + iI_0) + (R_0 + iI_0)I(R_0 - iI_0) + (R_0 - iI_0)(R -iI) I_0 \right] \\
                      &= \frac{2\beta V^2}{\omega}\left[ (R_0^2 - I_0^2)I + 2 R_0I_0 R \right] \, .
\end{align*}
Therefore,
\[
\tilde{G}_{\rho \rho} = \frac{2}{\beta \omega} \im \, \tilde{G}_{B \rho} .
\]

\section{Conventions and Fundamental Relationships}

We define the Fourier tranform of a function $f(t)$ as
\[
\mathcal{F}[f(t)](\omega) = f(\omega) = \int_{-\infty}^\infty e^{i \omega t} f(t) dt \, 
\]
with inverse transform
\[
\mathcal{F}^{-1}[f(\omega)](t) = 
    f(t) = \int_{-\infty}^\infty e^{-i \omega t} f(\omega) \frac{d\omega}{2\pi} \, .
\]
One of the tools we have used extensively is the Laplace transform:
\begin{align*}
\mathcal{L}[f(t)](z) &= \int_0^\infty e^{-zt} f(t) dt \, .
\end{align*}
We define the convolution of two functions $f$ and $g$ as:
\[
(f*g)(t) \equiv \int_0^t \, f(t - s) g(s) ds
\]
The Laplace transform of a convolution is then
\[
\mathcal{L}[(f*g)(t)] = \mathcal{L}[f(t)] \mathcal{L}[g(t)] \, . 
\]

In addition we have employed a ``discrete'' notation for various integral
kernels. This is convenient since we are only considering spherically
symmetric functions for which the notation corresponds to the dicretized 
form used in the numerical routines. For example, 
\[
C_{qk} f_k \equiv \int \frac{d^3 k}{(2 \pi)^3} C_{qk} f_k \,
\]
and
\[
\delta_{qk} \equiv (2 \pi)^3 \delta(\vec{q} - \vec{k}) \, .
\]

\end{appendices}

\bibliographystyle{unsrt}
\bibliography{FTSPD_HS}

\end{document}